\documentclass[journal,draftclsnofoot,onecolumn,12pt]{IEEEtran}

\ifCLASSINFOpdf
\else
\fi
\usepackage{amssymb}
\usepackage{gensymb}
\usepackage{graphicx}
\usepackage{amsmath}
\usepackage{float}
\usepackage{algorithm}
\usepackage{algpseudocode}

\usepackage{makecell}
\usepackage{cite}

\usepackage{geometry}

\usepackage{xcolor}
\usepackage{threeparttable}

\begin{document}
%
\title{Rate-Splitting Multiple Access for Multigateway Multibeam Satellite Systems with Feeder Link Interference}
%
%
%

\author{Zhi~Wen~Si,
        Longfei~Yin, 
        and~Bruno~Clerckx,~\IEEEmembership{Senior Member,~IEEE}
        
\thanks{Z. W. Si was with the Department of Electrical and Electronic Engineering, Imperial College, London SW7 2AZ, U.K. (email: zhiwen\_si@hotmail.com)}
\thanks{L. Yin and B. Clerckx are with the Communications and Signal Processing Group, Department of Electrical and Electronic Engineering, Imperial College, London SW7 2AZ, U.K. (email: longfei.yin17@imperial.ac.uk; b.clerckx@imperial.ac.uk)}


}

\maketitle
\vspace{-2.5em}
\begin{abstract}
This paper studies the precoder design problem of achieving max-min fairness (MMF) amongst users in multigateway multibeam satellite communication systems with feeder link interference. We propose a beamforming strategy based on a newly introduced transmission scheme known as rate-splitting multiple access (RSMA). RSMA relies on multi-antenna rate-splitting at the transmitter and successive interference cancellation (SIC) at the receivers, such that the intended message for a user is split into a common part and a private part and the interference is partially decoded and partially treated as noise. In this paper, we formulate the MMF problem subject to per-antenna power constraints at the satellite for the system with imperfect channel state information at the transmitter (CSIT). We also consider the case of two-stage precoding which is assisted by on-board processing (OBP) at the satellite. Numerical results obtained through simulations for RSMA and the conventional linear precoding method are compared. When RSMA is used, MMF rate gain is promised and this gain increases when OBP is used. RSMA is proven to be promising for multigateway multibeam satellite systems whereby there are various practical challenges such as feeder link interference, CSIT uncertainty, per-antenna power constraints, uneven user distribution per beam and frame-based processing.
\end{abstract}
%

%
\IEEEpeerreviewmaketitle

\section{Introduction}
%
%
%
%

 
With the recent launch of the fifth generation (5G) wireless technology, more devices are now able to connect to each other, forming an Internet of Things (IoT). These devices include not just cellular mobile phones but also other smart devices and sensors. This is made possible by implementing a multiple-input multiple-output (MIMO) system \cite{gesbert2007mimo} and expanding the available electromagnetic spectrum used for communication. Since the number of devices in the downlink has increased tremendously, interference between devices served by the same transmitter i.e. base station or satellite, becomes a significant problem when the transmitter is constantly reusing the entire available frequency spectrum to achieve a high throughput. In order to minimise this problem, advanced multiple access techniques and interference management strategies need to be designed and evaluated.

Different types of multiple access are being used to solve the problem, such as space-division multiple access (SDMA) and non-orthogonal multiple access (NOMA). SDMA involves the use of precoding methods at the transmitters to separate the users in the spatial domain. Beamforming is used to target the transmitter beam at the desired user and reduce the interference caused to other users. Residual interference is treated as noise at the receivers. On the other hand, NOMA involves the use of precoding methods at the transmitters to separate the users in the power domain. Messages intended for multiple users are precoded via superposition coding (SC) at the transmitter and the receiver of a single user will need to fully decode interfering streams and remove them via successive interference cancellation (SIC) before it decodes its own message.

Besides the conventional methods of using SDMA and NOMA, rate-splitting multiple access (RSMA) appeared as a promising technique to manage interference in multi-user multi-antenna networks \cite{clerckx2016rate, mao2018rate}. RSMA relies on multi-antenna rate-splitting (RS) at the transmitter and SIC at the receivers. Generally, the message for each user is split into a common part and a private part. For the simplest implementation of RS, known as 1-layer RS, all the common parts are combined and encoded as a single common stream whereas the private messages are encoded individually as private streams \cite{mao2018rate, joudeh2016sum}. At the receiver, the common stream is decoded and removed from the received signal via SIC. Then, the private stream intended for that user is decoded while treating all the other interfering streams as noise. Due to its flexibility in managing interference through common message decoding, RSMA serves as a general multiple access and boils down to SDMA, NOMA, orthogonal multiple access (OMA) and multicasting when allocating powers to the different types of message streams \cite{mao2018rate, clerckx2019rate}.

The utilisation of satellite communication systems is ever growing since they have a wide coverage and require minimal infrastructure on the ground. However, they have their own set of challenges to tackle such as on-board processing capability and line of sight problem with big path loss \cite{perez2019signal}. In order to achieve high throughput and scale the system capacity, aggressive frequency reuse with good interference management strategies are required \cite{vazquez2016precoding}. The channel model for a satellite system is fairly different as compared to a terrestrial system due to free space path loss and rain fading \cite{christopoulos2015multicast, zheng2012generic}.

In the forward link of a satellite system using a single gateway architecture, the user messages are transmitted from the gateway to the satellite via the feeder link, then transmitted from the satellite to the user terminals via the user link. Since the feeder link aggregates all satellite traffic, its bandwidth needs to be at least the total amount of bandwidth needed at the user link, which is the product of the number of feeds and the user link bandwidth \cite{joroughi2016precoding}. 

Due to the limited amount of bandwidth utilised for the feeder link and the high user demand, multiple gateways can be used in replace of the single gateway architecture, such that the feeder link spectrum can be reused through directive antennas \cite{joroughi2014multiple}. Another key benefit of this multiple gateway architecture when compared to the single gateway architecture is the reduction in the number of beams served by each gateway \cite{joroughi2014multiple, wang2019multicast} which would lead to a lower complexity. The total number of antenna feeds on the satellite are split among the gateways, leading to a smaller number of antenna feeds used by each gateway \cite{wang2019multicast}.

\begin{figure}
\centerline{\includegraphics[width=\columnwidth]{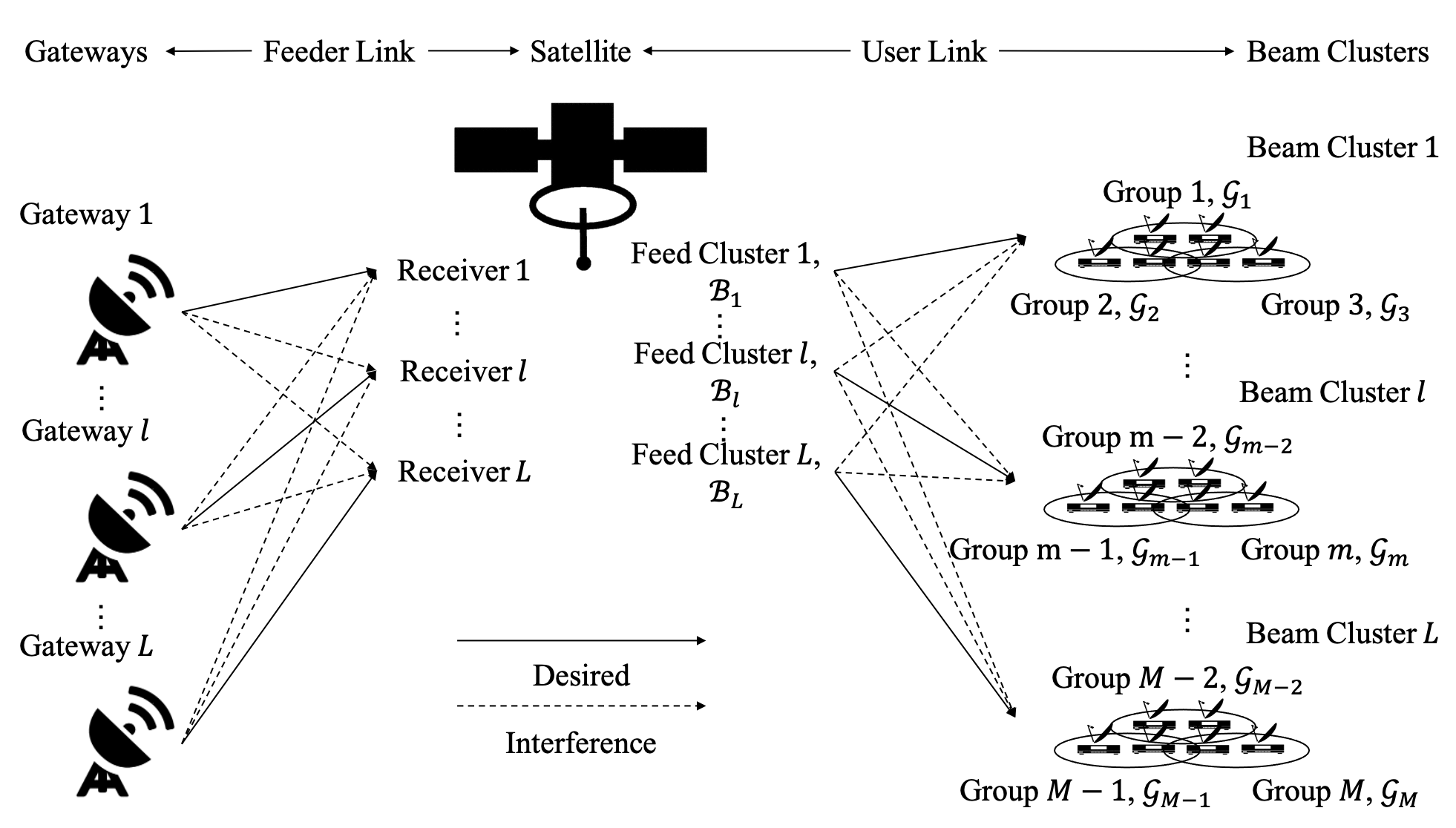}}
\caption{Architecture of multigateway multibeam satellite system.}
\label{fig_system}
\end{figure}

However, the use of multiple gateways introduces the problem of feeder link interference if there are pointing errors or the feeder link receivers are no longer calibrated correctly \cite{joroughi2016precoding}. This additional interference in the feeder link would cause the desired user signals to be corrupted by other unwanted user signals even before the multibeam interference in the user link, hence further deteriorating the user achievable rates. Also, the user link channel vector of a user terminal now varies based on the feed clusters since clusters of antenna feeds are managed by different gateways. Fig. \ref{fig_system} illustrates the multiple gateway architecture for a multibeam satellite system and the types of user signals available in the system.

In this paper, we investigate the novel transmission scheme of RSMA to mitigate the interference in the forward link of a multigateway multibeam satellite system and compare its performance with the conventional method of using linear precoding such as SDMA (denoted as NoRS in this paper). Imperfect channel state information at the transmitter (CSIT) will also be considered when designing the precoders to observe the robustness of the newly proposed transmission scheme.

\subsection{Related Works}
In the literature of multibeam satellite systems, the max sum rate multigroup multicast problem under both per antenna power and minimum rate constraints was introduced in \cite{christopoulos2015multicast}. Whereas in \cite{joroughi2017generalized}, a two-stage precoding design was proposed to handle the multibeam interference and then boost the lowest user's SINR in each beam for a multicast system. Since both papers only implemented a single gateway in their system models, no feeder link interference was taken into consideration. The use of multibeam satellites can also be found in an integrated terrestrial-satellite network whereby the two beamforming precoders were jointly optimised in \cite{zhu2018cooperative}.

In the literature of multigateway satellite systems, \cite{joroughi2014multiple} proposed a Zero Forcing (ZF) unicast precoder design while \cite{christopoulos2016multicast} proposed a max-min fairness (MMF) multicast precoder design to mitigate multibeam interference and cope with limited feeder link capacity. However, both papers neglected the effect of feeder link interference in their system models. As for \cite{wang2019multicast} and \cite{joroughi2016feeder}, both papers included the feeder link interference in their system models but assumed to have perfect CSIT at the gateways. This may not be that practical since perfect CSIT is usually hard to obtain.

Different with the above-mentioned works, the application of RSMA in multigateway multibeam satellite system considering feeder link interference and imperfect CSIT is investigated in this work. In the literature of RSMA, its benefits over other multiple accesses have been investigated and can be seen from many multi-antenna terrestrial systems, such as multiuser unicast systems with perfect CSIT \cite{mao2018rate, clerckx2019rate, alameer2019interference, zhang2019cooperative} and imperfect CSIT \cite{joudeh2016sum, hao2015rate, joudeh2016robust, piovano2017optimal, dai2016rate, mao2020beyond}, multigroup multicast systems with MMF problems \cite{joudeh2017rate, chen2020joint, yalcin2020rate} and energy efficiency maximisation problem \cite{tervo2018multigroup}, as well as superimposed unicast and multicast transmission \cite{mao2019rate}. Besides that, other applications of RSMA can also be seen in massive MIMO systems with residual transceiver hardware impairments \cite{papazafeiropoulos2017rate}, millimeter wave communications \cite{dai2017multiuser}, simultaneous wireless information and power transfer (SWIPT) networks \cite{mao2019wipt} and radar-communication (RadCom) systems \cite{xu2020rate}. Despite the different applications, RSMA outperforms the conventional methods in all of them. 

As for the use of RSMA in satellite systems, this can be seen in \cite{yin2020rate} whereby a per-feed power constrained MMF problem with different CSIT qualities is formulated for a multibeam satellite system using multigroup multicast beamforming. Also, \cite{vazquez2018rate} proposed a low complexity RS precoder design with perfect CSIT for a sum-rate optimisation problem by decoupling the design of the precoders for the common streams and the private streams. In \cite{caus2018exploratory}, RS is used in a two-beam satellite system and time division multiplexing (TDM) is considered within each beam. However, different from this paper, only a single gateway was considered and the feeder link was assumed to be noiseless in those three papers.

\subsection{Contributions}
The application of RSMA and underlying multi-antenna RS in a multigateway satellite communication system is analysed to observe its benefits over conventional methods in managing interference. The main contributions are summarised as follows.

\textit{First}, RSMA is for the first time applied to multibeam satellite systems with multiple gateways and feeder link interference. The RS strategy is implemented at each gateway and a centralised approach is used when designing the precoders. Global (perfect and imperfect) CSIT is available at the gateways. RS in multiple gateways leads to additional challenges whereby each gateway will need to compute a common stream from the group messages intended to the users served and include this additional stream in its transmit signal. At the user terminal, there are now additional interfering common streams that need to be treated as noise on top of the usual interfering private streams present in the received signal. Also, there are multiple satellite feed clusters transmitting the precoded symbol streams to the users. Therefore there are more than one user link channel vector used to calculate the SINR of the decoded streams for a specific user. This paper is an extension to multiple gateways of \cite{yin2020rate}, which focused on RSMA for multibeam satellite systems with a single gateway.
    
\textit{Second}, on-board processing (OBP), a promising capability for future satellite system is introduced and designed for the first time together with RS to mitigate the feeder link interference more effectively. By processing the received signals from the gateways at the satellite, the feeder link interference can be minimised before transmitting the signal to the user terminals. A two-stage precoding design is applied in this paper rather than a one-stage precoding design because there are two sources of interference and they can be handled separately and more effectively with the use of OBP. In the first stage, the first stage precoders and the OBP receiving filter matrices are optimised together using a minimum mean square error (MMSE) metric. As for the second stage, RS is applied when designing the second stage precoders. By doing so, the feeder link multigateway interference is controlled in the first stage and the user link multibeam interference is dealt with in the second stage. The two-stage precoding design in this paper contrasts with \cite{wang2019multicast} which demonstrated a leakage-based MMSE scheme for the first stage precoder design and applied a regular linear precoding in the second stage.
    
\textit{Third}, this is the first paper to investigate the feeder link channel under imperfect CSIT. Two different techniques are used to handle the feeder link channel estimates to minimise the error introduced to the actual channel. For the first technique, the ergodic MMF rate is calculated as the expectation of the average MMF rate over the variation of the channel estimate and Sample Average Approximation (SAA) is used to convert the stochastic problem into a deterministic approximated problem. A feeder link channel realisation set is formed by introducing random CSIT error back to the given channel estimate and the rate obtained for every realised channel is averaged. As for the second technique, the feeder link is treated the same as a MIMO channel with imperfect CSIT when carrying out OBP. A robust sum MSE minimisation problem is formulated by using a conditional MSE of the data stream for the given feeder link channel estimate. This contrasts with \cite{wang2019multicast, joroughi2016feeder} which focus on the gateways having perfect CSIT of the feeder link channel, therefore no channel error will need to be considered.
    
\textit{Fourth}, the MMF problem of RS subject to per-antenna power constraints at the satellite is formulated under both perfect and imperfect CSIT. This is the first work on the optimisation of RS for a multigateway multibeam satellite system with feeder link interference. One key difference between the MMF problems formulated in \cite{yin2020rate} and in this paper is that the power constraint is more complicated due to the impact of the feeder link interference and noise. The weighted MMSE (WMMSE) approach is used in this paper to make the formulated MMF problems convex since there is a rate-WMMSE relationship that leads to solving the problems using a low complexity algorithm based on alternating optimisation (AO).
    
\textit{Fifth}, the ergodic MMF rate is obtained through simulation and the numerical results obtained confirm that RSMA is capable of providing additional MMF rate gains in a multigateway architecture when compared to conventional methods. This is due to the fact that RSMA utilises the intra-gateway multibeam interference to its advantage, unlike NoRS that treats it completely as noise. When OBP is utilised in conjunction with RS, the gain achieved using RS over NoRS becomes very obvious. This is because the satellite transmitted signals are less corrupted by other gateways when OBP mitigates the feeder link interference, hence allowing RSMA to work on the user received signal even better. The RS strategy is proven to be promising for a multigateway multibeam satellite system whereby there are various practical challenges such as feeder link interference, CSIT uncertainty, per-antenna power constraints, uneven user distribution per beam and frame-based processing.

\subsection{Organisation}
The remainder of the paper is organised as follows. Section II describes the system model. In Section III, the MMF problem is formulated for a noisy feeder link with interference. The two-stage precoding with on-board processing is also introduced. The optimisation algorithms are described in Section IV. Finally, Section V presents the simulation results and Section VI concludes the paper.

\subsection{Notation}
The following notations are used in this paper. $a$ is a scalar, $\mathbf{a}$ is a column vector, $\mathbf{A}$ is a matrix and $\mathcal{A}$ is a set. $\left( \cdot \right)^T$, $\left( \cdot \right)^H$, $\left| \cdot \right|$, $\left\| \cdot \right\|$, $\left| \mathcal{A} \right|$, $\mathbb{E}\left\{ \cdot \right\}$, $\mathcal{R}\left\{ \cdot \right\}$ and $\mathrm{Tr}\left\{ \mathbf{A} \right\}$ correspond to transpose, Hermitian transpose, absolute value, Euclidean norm, cardinality of a set, expectation, real part of a complex number and trace of a matrix respectively. $\mathbb{R}$ and $\mathbb{C}$ denote real and complex space. $\mathbf{I}$ is an identity matrix, $\mathbf{D}_n$ is a zero square matrix but with 1 as the $n$-th diagonal element and $\mathbf{1}$ is a vector with all-ones.

\section{System Model}
For this paper, the satellite system operates on the Ka-band of the electromagnetic spectrum and the satellite has a geostationary orbit (GEO) \cite{yin2020rate}, serving multiple terrestrial user terminals as shown in Fig. \ref{fig_system}. There are $N$ antenna feeds on the satellite and the set of antenna feeds is denoted as $\mathcal{N} = \left\{ 1,\ldots,N \right\}$. It is assumed that each feed produces a single beam. Based on the satellite standard DVB-S2X \cite{dvb}, multiple users within the same beam communicate the same coded frame during a transmission period. Each beam covers a single multicast group, thus generating $M = N$ groups in the system.

There are $K$ users in the system and each of them is equipped with a single antenna, forming a multigroup multicast downlink multiple-input single-output (MISO) system. The set of all users is denoted by $\mathcal{K} = \left\{ 1,\ldots,K \right\}$. These users locate in the $M \left( 1 \leq M \leq K \right)$ groups such that users in the same group receive the same multicast message. These messages are considered to be independent amongst the different groups \cite{yin2020rate}. Each group consists of a subset of users, denoted by $\mathcal{G}_m$ for all $m \in \mathcal{M} = \left\{ 1,\ldots,M \right\}$. When grouping the users, the mapping $\mu \colon \mathcal{K} \rightarrow \mathcal{M}$ is used such that $\mu \left( k \right) = m$, $\forall k \in \mathcal{G}_m$ \cite{joudeh2017rate}. Every user is part of a single group only, i.e. $\mathcal{G}_i \cap \mathcal{G}_j = \emptyset$ if $i \neq j$, $\forall i,j \in \mathcal{M}$. Since there are no groups sharing the same user, we have $\sum_{m\in\mathcal{M}} G_m = K$ if the size of group $m$ is denoted by $G_m = \left| \mathcal{G}_m \right|$. Without loss of generality, the group sizes are arranged in an ascending order, i.e. $G_1 \leq G_2 \leq \cdots \leq G_M$ \cite{joudeh2017rate}.

When $L > 1$ gateways are used, $B_l \left( 1 \leq B_l < N \right)$ satellite antenna feeds are used by the same gateway $l$, forming a feed cluster \cite{wang2019multicast}. The set of gateways used is denoted by $\mathcal{L} = \left\{ 1,\ldots,L \right\}$ while the set of antenna feeds in each cluster is denoted by $\mathcal{B}_l$ such that $\mathcal{B}_l \subset \mathcal{N}$. No antenna feed is shared among gateways, hence $\mathcal{B}_i \cap \mathcal{B}_j = \emptyset$ if $i \neq j$, $\forall i,j \in \mathcal{L}$. When assigning a group to be served by a gateway, the mapping $\lambda \colon \mathcal{M} \rightarrow \mathcal{L}$ is used such that $\lambda \left(m \right) = l$, $\forall m \in \mathcal{B}_l$ means that gateway $l$ serves all groups $m \in \mathcal{B}_l$. Without loss of generality, the gateways are arranged in an ascending order based on the number of groups served by each of them, i.e. $B_1 \leq B_2 \leq \cdots \leq B_L$.

\subsection{Channel Model}
\subsubsection{User Link}
The user link channel vector of dimension $B_l \times 1$ between feed cluster $l$ and user $k$ can be written as a Hadamard product of two vectors
\begin{equation}
    \mathbf{h}_{l,k} = \mathbf{b}_{l,k} \circ \mathbf{q}_{l,k},
\end{equation}
where $\mathbf{b}_{l,k} \in \mathbb{R}^{B_l \times 1}$ represents the free space loss, the receiver antenna gain and the satellite beam gain and $\mathbf{q}_{l,k} \in \mathbb{C}^{B_l \times 1}$ represents the phase and the atmospheric fading due to rain.

In order to formulate the vector $\mathbf{b}_{l,k}$, following \cite{esa}, the $n$-th element of the vector is calculated using
\begin{equation}
    b_{n,l,k} = \frac{\sqrt{G_R G_{n,l,k}}}{4\pi \frac{d_k}{\lambda} \sqrt{\kappa T_{sys} B_w}},
\end{equation}
where $G_R$ is the receiver antenna gain, $d_k$ is the distance between the satellite and user $k$, $\lambda$ is the carrier wavelength, $\kappa$ is the Boltzmann constant, $T_{sys}$ is the system noise temperature in kelvin and $B_w$ is the user link bandwidth. 

According to \cite{caini1992spectrum}, $G_{n,l,k}$ is the satellite beam gain and can be calculated as
\begin{equation}
    G_{n,l,k} = G_{max} \left[ \frac{J_1 \left( u_{n,l,k} \right)}{2u_{n,l,k}} + 36\frac{J_3 \left( u_{n,l,k} \right)}{u^3_{n,l,k}} \right]^2,
\end{equation}
where $G_{max}$ is the maximum beam gain observed at the beam centre and $J_\alpha \left( x \right)$ is the Bessel function of the first kind with order $\alpha$. The parameter $u_{n,l,k} = 2.07123 \sin \left( \theta_{n,l,k} \right) / \sin \left( \theta _{\mathrm{3dB}} \right)$ whereby $\theta_{n,l,k}$ is the angle between the $n$-th beam centre of feed cluster $l$ and user $k$ while $\theta _{\mathrm{3dB}}$ is the angle for which a 3 dB power loss occurs when compared to the beam centre.

As for the formulation of vector $\mathbf{q}_{l,k}$, following \cite{itu}, the $n$-th element of the vector is calculated using
\begin{equation}
    q_{n,l,k} = \chi_k^{-\frac{1}{2}} e^{-j\phi_k}, \label{equ_rain_fading}
\end{equation}
where the rain attenuation gain in dB, $\chi_{k,dB} = 20\log_{10} \left( \chi_k \right)$, follows a lognormal distribution, i.e. $\ln \left( \chi_{k,dB} \right) \sim \mathcal{N} \left( \mu,\sigma^2 \right)$, while the phase $\phi_k$ follows a uniform distribution between $0$ and $2\pi$. 

According to (\ref{equ_rain_fading}), both the rain attenuation gain and phase remain the same regardless of the antenna feeds since they only change when the user changes. This is due to the fact that the spacing between the antenna feeds is relatively small when compared to the long signal propagation path distance \cite{zheng2012generic}.

\subsubsection{Feeder Link}
The feeder link channel between gateway $l$ and feeder link receiver $i$ is denoted as $\mathbf{F}_{i,l}$. This channel has a dimension of $B_l \times B_l$ such that they match the number of antenna feeds in the feed cluster managed by gateway $l$. The feeder link channel model is kept general throughout the main body of this paper and a specific model will be described in Section \ref{sec_simulation_result} when the simulation results are presented.

\subsection{Signal Model}
\label{sec_transmission_scheme}
For the multigroup multicasting scenario in beam cluster $l$, group message $W_m$ is intended for the users in $\mathcal{G}_m$, $\forall m \in \mathcal{B}_l$. When applying the RS strategy, each of these messages are split into a common part and a private part, i.e. $W_m \rightarrow \left\{ W_{m,c}, W_{m,p} \right\}$. All the common parts are combined together and encoded into a common stream, i.e. $\left\{ W_{m,c} \mid m \in \mathcal{B}_l \right\} \rightarrow s_{c,l}$. As for the private parts, they are encoded individually into private streams independently, i.e. $\left\{ W_{m,p} \right\} \rightarrow s_m$, $\forall m \in \mathcal{B}_l$. Gateway $l$ now has a vector of symbol streams, $\mathbf{s}_l \in \mathbb{C}^{\left( B_l+1 \right) \times 1}$, which is composed of $s_{c,l}$ and $s_m$, $\forall m \in \mathcal{B}_l$. It is assumed that $\mathbb{E} \left\{ \mathbf{s}_l\mathbf{s}_l^H \right\} = \mathbf{I}$. A linear precoding matrix $\mathbf{P}_l \in \mathbb{C}^{B_l \times \left( B_l+1 \right)}$ is used to map the symbol streams to the transmit antennas and the resultant transmit signal $\mathbf{x}_l$ for gateway $l$ is written as
\begin{equation}
    \mathbf{x}_l = \mathbf{P}_l\mathbf{s}_l = \mathbf{p}_{c,l}s_{c,l} + \sum_{m \in \mathcal{B}_l} \mathbf{p}_m s_m,
\end{equation}
where $\mathbf{p}_{c,l} \in \mathbb{C}^{B_l \times 1}$ is the precoder for the common stream and $\mathbf{p}_m \in \mathbb{C}^{B_l \times 1}$ is the precoder for the private stream. At each gateway, the sum power constraint is employed to limit the amount of power used to transmit the signal to the satellite. This power constraint is written as

\begin{equation}
    \mathbf{p}^H_{c,l} \mathbf{p}_{c,l} + \sum_{m \in \mathcal{B}_l} \mathbf{p}^H_m \mathbf{p}_m \leq P_l,\ \forall l \in \mathcal{L},
\end{equation}
where $P_l$ is the power limit for gateway $l$.

At the satellite, there are $L$ feeder link receivers corresponding to the $L$ gateways. The received signal at feeder link receiver $l$ is a combination of signals transmitted from all gateways and can be written as
\begin{equation}
    \mathbf{r}_l = \sum_{j=1}^L \mathbf{r}_{l,j} + \mathbf{n}_l = \sum_{j=1}^L \mathbf{F}_{l,j} \mathbf{x}_j + \mathbf{n}_l,
\end{equation}
where $\mathbf{r}_{l,j}$, $\forall j \neq l$ is the interfering signal and $\mathbf{n}_l$ is the additive white Gaussian noise (AWGN). We assume that $\mathbb{E} \left\{ \mathbf{n}_l \mathbf{n}_l^H \right\} = \sigma_n^2 \mathbf{I}$.

Before passing the received signals to the HPA and the feeds, an OBP receiving filter matrix $\mathbf{R}_l \in \mathbb{C}^{B_l \times B_l}$ based on the type of payload is used at feeder link receiver $l$ to modify the received signal such that $\overline{\mathbf{r}}_l = \mathbf{R}_l \mathbf{r}_l$. The transmit power constraint at the satellite is the per-antenna power constraint since each antenna has a dedicated high power amplifier (HPA) that may be operating close to saturation. These amplifiers do not allow for energy resources to be shared freely amongst the antennas \cite{perez2019signal}. The satellite transmit power constraint is written as
\begin{align}
    & \sum_{j=1}^L \mathbf{p}^H_{c,j} \mathbf{F}^H_{l,j} \mathbf{R}^H_l \mathbf{D}_{f \left( n,\mathcal{B}_l \right)} \mathbf{R}_l \mathbf{F}_{l,j} \mathbf{p}_{c,j} + \sum_{m \in \mathcal{M}} \mathbf{p}^H_m \mathbf{F}^H_{l,\lambda \left( m \right)} \mathbf{R}^H_l \mathbf{D}_{f \left( n,\mathcal{B}_l \right)} \mathbf{R}_l \mathbf{F}_{l,\lambda \left( m \right)} \mathbf{p}_m \nonumber \\
    & + \sigma^2_n \left[ \mathbf{R}_l \mathbf{R}^H_l \right]_{f \left( n,\mathcal{B}_l \right)f \left( n,\mathcal{B}_l \right)} \leq P_n,\ \forall n \in \mathcal{B}_l, l \in \mathcal{L},
\end{align}
where $P_n$ is the allocated power for antenna feed $n$ and the function $f \left( n,\mathcal{B}_l \right)$ returns the index of antenna feed $n$ in the antenna feed set $\mathcal{B}_l$. The transmit power constraint at the satellite is more stringent than the transmit power constraint at the gateways because power is more readily available for the gateways on Earth and there is no flexible power sharing available at the satellite. Consequently, we assume that there is enough power budget at every gateway and we ignore the transmit power constraint at all gateways.

At user $k$, the received signal $y_k = \sum_{l=1}^L \mathbf{h}^H_{l,k} \overline{\mathbf{r}}_l + n_k$, $\forall k \in \mathcal{K}$. According to the equation, $\mathbf{h}_{l,k} \in \mathbb{C}^{B_l \times 1}$ is the user link channel vector between feed cluster $l$ and user $k$, and $n_k \sim \mathcal{CN} \left( 0,\sigma^2_{n,k} \right)$ is the AWGN. This noise is independent and identically distributed (i.i.d.) across all users and it is assumed that $\sigma^2_{n,k} = \sigma^2_{n}$ without loss of generality \cite{yin2020rate}.

The received signal of user $k$ can be expanded as follows
\begin{subequations}
\begin{align}
    y_k & = \sum_{l=1}^L \mathbf{h}^H_{l,k} \mathbf{R}_l \mathbf{r}_l + n_k \\
    & = \sum_{l=1}^L \sum_{j=1}^L \mathbf{h}^H_{l,k} \mathbf{R}_l \mathbf{F}_{l,j} \mathbf{p}_{c,j}s_{c,j} + \sum_{l=1}^L \sum_{j=1}^L \sum_{m \in \mathcal{B}_j} \mathbf{h}^H_{l,k} \mathbf{R}_l \mathbf{F}_{l,j} \mathbf{p}_m s_m + \sum_{l=1}^L \mathbf{h}^H_{l,k} \mathbf{R}_l \mathbf{n}_l + n_k \\
    & = \sum_{j=1}^L \overline{\mathbf{h}}^H_{j,k} \mathbf{p}_{c,j}s_{c,j} + \sum_{j=1}^L \sum_{m \in \mathcal{B}_j} \overline{\mathbf{h}}^H_{j,k} \mathbf{p}_m s_m + \overline{n}_k \label{equ_received_signal} \\
    & = \overline{\mathbf{h}}^H_{\lambda \left[ \mu \left( k \right) \right],k} \mathbf{p}_{c,\lambda \left[ \mu \left( k \right) \right]} s_{c,\lambda \left[ \mu \left( k \right) \right]} + \overline{\mathbf{h}}^H_{\lambda \left[ \mu \left( k \right) \right],k} \mathbf{p}_{\mu \left( k \right)} s_{\mu \left( k \right)} + \underbrace{\sum_{m \in \mathcal{B}_{\lambda \left[ \mu \left( k \right) \right]} \setminus \mu \left( k \right)} \overline{\mathbf{h}}^H_{\lambda \left[ \mu \left( k \right) \right],k} \mathbf{p}_m s_m}_\mathrm{intra-gateway interference} \nonumber \\
    & + \underbrace{\sum_{j \in \mathcal{L} \setminus \lambda \left[ \mu \left( k \right) \right]} \left[\overline{\mathbf{h}}^H_{j,k} \mathbf{p}_{c,j} s_{c,j} + \sum_{m \in \mathcal{B}_j} \overline{\mathbf{h}}^H_{j,k} \mathbf{p}_m s_m \right]}_\mathrm{inter-gateway interference} + \overline{n}_k,
\end{align}
\end{subequations}
where the effective channel between gateway $j$ and user $k$ is $\overline{\mathbf{h}}^H_{j,k} = \sum_{l=1}^L \mathbf{h}^H_{l,k} \mathbf{R}_l \mathbf{F}_{l,j}$ and the effective noise is $\overline{n}_k = \sum_{l=1}^L \mathbf{h}^H_{l,k} \mathbf{R}_l \mathbf{n}_l + n_k$ \cite{wang2019multicast}.

For user $k$, the desired common stream is $s_{c,\lambda \left[ \mu \left( k \right) \right]}$. This stream will be decoded first and all the other streams will be treated as noise. The signal-to-interference-plus-noise ratio (SINR) of decoding $s_{c,\lambda \left[ \mu \left( k \right) \right]}$ at user $k$ is given by
\begin{equation}
    \gamma_{c,k} = \frac{\left|\overline{\mathbf{h}}^H_{\lambda \left[ \mu \left( k \right) \right],k} \mathbf{p}_{c,\lambda \left[ \mu \left( k \right) \right]}\right|^2}{I_c + \overline{\sigma}^2_{n,k}},
\end{equation}
where we denote $I_c = \left|\overline{\mathbf{h}}^H_{\lambda \left[ \mu \left( k \right) \right],k} \mathbf{p}_{\mu \left( k \right)}\right|^2 + \sum_{m \in \mathcal{B}_{\lambda \left[ \mu \left( k \right) \right]} \setminus \mu \left( k \right)} \left|\overline{\mathbf{h}}^H_{\lambda \left[ \mu \left( k \right) \right],k} \mathbf{p}_m\right|^2 + \sum_{j \in \mathcal{L} \setminus \lambda \left[ \mu \left( k \right) \right]} \Bigg[\left|\overline{\mathbf{h}}^H_{j,k} \mathbf{p}_{c,j}\right|^2$ $+ \sum_{m \in \mathcal{B}_j}\left|\overline{\mathbf{h}}^H_{j,k} \mathbf{p}_m\right|^2\Bigg]$ and $\overline{\sigma}^2_{n,k} = \sigma^2_n \sum_{l=1}^L \left|\mathbf{h}^H_{l,k} \mathbf{R}_l \right|^2 + \sigma^2_n$.

The achievable rate by user $k$ for the desired common stream is $R_{c,k} = \log_2 \left( 1+\gamma_{c,k} \right)$. In order to ensure that $s_{c,\lambda \left[ \mu \left( k \right) \right]}$ can be decoded by all users in the groups within beam cluster $l = \lambda \left[ \mu \left( k \right) \right]$, the achievable rate by the weakest user in that beam cluster is the common rate at which $s_{c,\lambda \left[ \mu \left( k \right) \right]}$ is communicated, i.e. $R_{c,l} = \min_{k \in \mathcal{G}_m, m \in \mathcal{B}_l} R_{c,k}$.

Since $s_{c,\lambda \left[ \mu \left( k \right) \right]}$ is intended for all the groups in beam cluster $l = \lambda \left[ \mu \left( k \right) \right]$, the common rate is split into portions such that $R_{c,l} = \sum_{m \in \mathcal{B}_l} C_m$. After removing the desired common stream from the received signal using SIC, the desired private stream $s_{\mu \left( k \right)}$ for user $k$ is then decoded and all the other interfering streams are treated as noise. The SINR of decoding $s_{\mu \left( k \right)}$ at user $k$ is given by
\begin{equation}
    \gamma_k = \frac{\left|\overline{\mathbf{h}}^H_{\lambda \left[ \mu \left( k \right) \right],k} \mathbf{p}_{\mu \left( k \right)}\right|^2}{I_c - \left|\overline{\mathbf{h}}^H_{\lambda \left[ \mu \left( k \right) \right],k} \mathbf{p}_{\mu \left( k \right)}\right|^2 + \overline{\sigma}^2_{n,k}}.
\end{equation}
    
The achievable rate by user $k$ for the desired private stream is $R_k = \log_2 \left( 1+\gamma_k \right)$. However, since $s_{\mu \left( k \right)}$ also needs to be decoded by all users within group $m = \mu \left( k \right)$ with guarantee, the rate at which $s_{\mu \left( k \right)}$ is communicated is therefore the minimum achievable rate by all users in that group, i.e. $r_m = \min_{i \in \mathcal{G}_m} R_i$. 
As a result of decoding the two desired streams, the overall $m$-th group-rate is written as
\begin{equation}
    r_{g,m}^{RS} = C_m + r_m.
\end{equation}

As for NoRS at gateway $l$, the messages are encoded into private streams directly, i.e. $\left\{ W_m \right\} \rightarrow s_m$, $\forall m \in \mathcal{B}_l$. The vector of symbol streams, $\mathbf{s}_l \in \mathbb{C}^{\left( B_l \right) \times 1}$ is composed of $s_m$, $\forall m \in \mathcal{B}_l$ and $\mathbb{E} \left\{ \mathbf{s}_l\mathbf{s}_l^H \right\} = \mathbf{I}$. User $k$ will only need to decode its desired stream $s_{\mu \left( k \right)}$ and treat all the other interfering streams as noise. NoRS can be regarded as a special case of RS by turning off its common stream and allocating all the transmit power to its private streams. No SIC is required at receivers. The SINR of decoding $s_{\mu \left( k \right)}$ at user $k$ is given by
\begin{equation}
    \gamma_k = \frac{\left|\overline{\mathbf{h}}^H_{\lambda \left[ \mu \left( k \right) \right],k} \mathbf{p}_{\mu \left( k \right)}\right|^2}{I + \overline{\sigma}^2_{n,k}},
\end{equation}
where we denote $I = \sum_{m \in \mathcal{B}_{\lambda \left[ \mu \left( k \right) \right]} \setminus \mu \left( k \right)} \left|\overline{\mathbf{h}}^H_{\lambda \left[ \mu \left( k \right) \right],k} \mathbf{p}_m\right|^2 + \sum_{j \in \mathcal{L} \setminus \lambda \left[ \mu \left( k \right) \right]} \sum_{m \in \mathcal{B}_j}\left|\overline{\mathbf{h}}^H_{j,k} \mathbf{p}_m\right|^2$.

Once again, the achievable rate by user $k$ for the desired stream is $R_k = \log_2 \left( 1+\gamma_k \right)$ and the $m$-th group-rate is written as
\begin{equation}
    r_{g,m}^{NoRS} = \min_{i \in \mathcal{G}_m} R_i = r_m.
\end{equation}

\subsection{Channel State Information}
We assume that the receiver has perfect channel state information, i.e. perfect CSIR. When the CSI is estimated at the receiver, quantised and then fed back to the transmitter, CSIT is a realistic practical issue.

When imperfect CSIT is considered, the user link composite channel $\mathbf{H} = \widehat{\mathbf{H}} + \widetilde{\mathbf{H}}$ whereby $\widehat{\mathbf{H}} = \big[ \widehat{\mathbf{h}}_1,\ldots,\widehat{\mathbf{h}}_K \big]$ is the composite channel of the estimated channel vectors and $\widetilde{\mathbf{H}} = \big[ \widetilde{\mathbf{h}}_1,\ldots,\widetilde{\mathbf{h}}_K \big]$ is the composite channel of the channel error vectors \cite{joudeh2016sum}. The CSIT error variance is calculated as $\sigma_{e,k}^2 = \mathbb{E}_{\widetilde{\mathbf{h}}_k} \big\{\big\| \widetilde{\mathbf{h}}_k \big\|^2\big\}$ \cite{joudeh2016sum}. For simplicity, the error power for every user is assumed to be the same, i.e. $\sigma_{e,k}^2 = \sigma_e^2$, and it is allowed to scale as $\mathit{O} \left( P^{-\alpha} \right)$ \cite{yang2013degrees}. The scaling factor $\alpha$ quantifies the CSIT quality and it takes the value from the range $0$ to $\infty$ \cite{joudeh2016sum}. The same concept is also applied to the feeder link channel such that $\mathbf{F} = \widehat{\mathbf{F}} + \widetilde{\mathbf{F}}$.

If $\alpha = 0$, the CSIT quality remains constant regardless of the SNR. This can be an example of a case when the number of feedback bits used is constant \cite{joudeh2016sum}. However, if $\alpha \rightarrow \infty$, perfect CSIT is achieved since $\sigma_e^2 \rightarrow 0$. This can be an example of a case when the number of feedback bits used is infinite \cite{yin2020rate}. As for a finite $\alpha > 0$, the CSIT quality improves when the SNR increases since the error decays more. This can be an example of a case when the number of feedback bits used is increasing \cite{joudeh2016sum}. The range of values $\alpha$ can take is truncated such that $\alpha \in \left[ 0,1 \right]$ because $\alpha = 1$ corresponded to perfect CSIT from a DoF perspective \cite{yang2013degrees}.

\textit{Remark 1:} Imperfect CSIT should not be considered for the user link only because imperfect CSIT of the feeder link would also result in inaccurate precoder design. Therefore, imperfect feeder link CSIT should be considered in the system model as well to determine the practicality of the transmission schemes.

\section{Problem Formulation}
When RS is implemented and imperfect CSIT is considered, the gateways are unable to predict the instantaneous rate for the common and private streams. However, they are able to obtain the stochastic Average Rate (AR) for all streams, which are short-term measures that capture the expected system performance over the CSIT error distribution for a given channel state estimate \cite{joudeh2016sum}. The ARs for user $k$ are defined as $\overline{R}_{c,k} \big( \widehat{\overline{\mathbf{H}}} \big) = \mathbb{E}_{\overline{\mathbf{H}} \mid \widehat{\overline{\mathbf{H}}}} \big\{ R_{c,k} \big( \overline{\mathbf{H}}, \widehat{\overline{\mathbf{H}}} \big) \mid \widehat{\overline{\mathbf{H}}} \big\}$ and $\overline{R}_k \big( \widehat{\overline{\mathbf{H}}} \big) = \mathbb{E}_{\overline{\mathbf{H}} \mid \widehat{\overline{\mathbf{H}}}} \big\{ R_k \big( \overline{\mathbf{H}},\widehat{\overline{\mathbf{H}}} \big) \mid \widehat{\overline{\mathbf{H}}} \big\}$, whereby $\overline{\mathbf{H}}$ is the composite effective channel between the gateways and the users, while $\widehat{\overline{\mathbf{H}}}$ is its estimate.

The MMF Ergodic Rate (ER) is used as the performance criteria for all the transmission schemes discussed in this paper. Note that the ER is a long-term measure that captures the expected system performance over all channel states. Based on the law of total expectation, the ERs for user $k$ can be expressed as $\mathbb{E}_{\big\{ \overline{\mathbf{H}},\widehat{\overline{\mathbf{H}}} \big\}} \big\{ R_{c,k} \big( \overline{\mathbf{H}},\widehat{\overline{\mathbf{H}}} \big) \big\} = \mathbb{E}_{\widehat{\overline{\mathbf{H}}}} \big\{  \mathbb{E}_{\overline{\mathbf{H}} \mid \widehat{\overline{\mathbf{H}}}} \big\{ R_{c,k} \big( \overline{\mathbf{H}},\widehat{\overline{\mathbf{H}}} \big) \mid \widehat{\overline{\mathbf{H}}} \big\} \big\} = \mathbb{E}_{\widehat{\overline{\mathbf{H}}}} \left\{\overline{R}_{c,k} \left(\widehat{\overline{\mathbf{H}}}\right)\right\}$ and $\mathbb{E}_{\big\{ \overline{\mathbf{H}},\widehat{\overline{\mathbf{H}}} \big\}} \big\{ R_k \big( \overline{\mathbf{H}},\widehat{\overline{\mathbf{H}}} \big) \big\} = \mathbb{E}_{\widehat{\overline{\mathbf{H}}}} \big\{  \mathbb{E}_{\overline{\mathbf{H}} \mid \widehat{\overline{\mathbf{H}}}} \big\{ R_k \big( \overline{\mathbf{H}},\widehat{\overline{\mathbf{H}}} \big) \mid \widehat{\overline{\mathbf{H}}} \big\} \big\} = \mathbb{E}_{\widehat{\overline{\mathbf{H}}}} \left\{\overline{R}_k \left(\widehat{\overline{\mathbf{H}}}\right)\right\}$.

It can be seen that the ER for each (common or private) stream is calculated as the expectation of the AR over the variation of $\widehat{\overline{\mathbf{H}}}$ \cite{joudeh2016sum}. Therefore, a stochastic MMF AR problem for a given effective channel estimate needs to be formulated and solved. 

In order to solve the problem, a deterministic approximation of the problem is formed by adopting the SAA method. For a given user link channel estimate $\widehat{\mathbf{H}}$, a set of $S$ i.i.d. realisations drawn from a conditional distribution with density $f_{\mathbf{H} \mid \widehat{\mathbf{H}}} \big(\mathbf{H} \mid \widehat{\mathbf{H}}\big)$ is constructed \cite{joudeh2016sum}. The set of sample index is denoted by $\mathfrak{S} = \left\{ 1,\ldots,S \right\}$ and the realisation set $\mathbb{H}^{ \left( S \right)}$ is given by
\begin{equation}
    \mathbb{H}^{\left( S \right)} = \left\{\mathbf{H}^{ \left( s \right)} = \widehat{\mathbf{H}} + \widetilde{\mathbf{H}}^{\left( s \right)} \mid \widehat{\mathbf{H}},\ s \in \mathfrak{S}\right\},
\end{equation}
where $\mathbf{H}^{ \left( s \right)}$ is the $s$-th conditional realisation of the user link channel and $\widetilde{\mathbf{H}}^{ \left( s \right)}$ is drawn from the CSIT error distribution of $\mathcal{CN} \left( 0,\sigma_e^2 \right)$.

A realisation set $\mathbb{F}^{ \left( S \right)}$ is also created for a given feeder link channel estimate $\widehat{\mathbf{F}}$. The feeder link and user link channel realisation sets are then used to create a realisation set for the effective channel between the gateways and the users $\overline{\mathbb{H}}^{ \left( S \right)}$. These channel realisations are available at the gateways and used to approximate the AR for each common stream and each private stream by averaging the rates achieved over the $S$ channel realisations \cite{joudeh2016sum}. When $S \rightarrow \infty$, due to the strong law of large numbers, the ARs of user $k$ are given by
\begin{subequations}
\begin{gather}
    \overline{R}_{c,k} = \lim_{S \rightarrow \infty} \overline{R}_{c,k}^{ \left( S \right)} = \lim_{S \rightarrow \infty} \frac{1}{S} \sum_{s=1}^S R_{c,k}\left(\mathbf{H}^{ \left( s \right)}, \mathbf{F}^{ \left( s \right)}\right), \\
    \overline{R}_k = \lim_{S \rightarrow \infty} \overline{R}_k^{ \left( S \right)} = \lim_{S \rightarrow \infty} \frac{1}{S} \sum_{s=1}^S R_k\left(\mathbf{H}^{ \left( s \right)}, \mathbf{F}^{ \left( s \right)}\right).
\end{gather}
\end{subequations}

\subsection{Noisy Feeder Link with Interference}
\label{sec_feeder}
In this subsection, the precoder design problem is formulated for the transmission schemes described in Section \ref{sec_transmission_scheme} with no on-board processing carried out at the satellite. The received signals at the satellite are passed straight through to the HPA and then routed to the antenna feeds, therefore the OBP receiving filter matrices are merely identity matrices, i.e. $\mathbf{R}_l = \mathbf{I}$, $\forall l \in \mathcal{L}$. For this paper, the MMF metric is used whereby the lowest group-rate is maximised.

Subject to per-antenna power constraints at the satellite, the deterministic approximation of the MMF problem using the RS strategy at multiple gateways under imperfect CSIT is formulated as

$\mathcal{F}^{\left( S \right)}_{RS_{feeder}} \left( P \right)$:
\begin{subequations}
\begin{align}
\max_{\overline{\mathbf{c}},\mathbf{P}} \quad & \min_{m \in \mathcal{M}} \left(\overline{C}_m + \overline{r}_m\right) \\
s.t. \quad & \overline{R}_{c,k}^{\left( S \right)} \geq \sum_{i \in \mathcal{B}_l} \overline{C}_i,\ \forall k \in \mathcal{G}_m, m \in \mathcal{B}_l, l \in \mathcal{L}, \\
& \overline{C}_m \geq 0,\ \forall m \in \mathcal{M}, \label{equ_common_rate_portion}\\
& \sum_{j=1}^L \mathbf{p}^H_{c,j} \widehat{\mathbf{F}}^H_{l,j} \mathbf{D}_{f \left( n,\mathcal{B}_l \right)} \widehat{\mathbf{F}}_{l,j} \mathbf{p}_{c,j} + \sum_{m \in \mathcal{M}} \mathbf{p}^H_m \widehat{\mathbf{F}}^H_{l,\lambda \left( m \right)} \mathbf{D}_{f \left( n,\mathcal{B}_l \right)} \widehat{\mathbf{F}}_{l,\lambda \left( m \right)} \mathbf{p}_m \nonumber \\
& + \sigma^2_n \leq P_n,\ \forall n \in \mathcal{B}_l, l \in \mathcal{L}.
\end{align}
\end{subequations}

The average common rate portions $\overline{\mathbf{c}} = \left[ \overline{C}_1, \ldots, \overline{C}_M \right]$ and precoders $\mathbf{P} = \big[ \mathbf{p}_{c,1}, \ldots, \mathbf{p}_{c,L}, \mathbf{p}_1, \ldots,$ $\mathbf{p}_M \big]$ are jointly optimised when the problem is solved. Constraint (\ref{equ_common_rate_portion}) ensures that group $m$ does not have a negative rate portion when splitting the common rate $R_{c,\lambda \left( m \right)}$.

The problem is formulated under imperfect CSIT because it is more general than perfect CSIT. If perfect CSIT is available at the gateways, the channel estimates would turn into the actual channels and the ARs would turn into the instantaneous rates when formulating the MMF problem. The problem formulated under perfect CSIT can be seen as a special case of $\mathcal{F}^{\left( S \right)}_{RS_{feeder}} \left( P \right)$.

As for NoRS at the gateways, the MMF problem formulated is also a special case of $\mathcal{F}^{\left( S \right)}_{RS_{feeder}} \left( P \right)$ whereby $\overline{\mathbf{c}} = 0$ and $\| \mathbf{p}_{c,l} \|^2 = 0$, $\forall l \in \mathcal{L}$. Moving forward, the problem formulated for the RS strategy will be discussed since NoRS is a subset of RS.

\subsection{On-Board Processing}
\label{sec_obp}
The possibility of on-board processing is looked into to minimise the feeder link interference before passing the received signal to the satellite feeds for user link transmission. By processing the received signals at the satellite, a two-stage precoding design \cite{wang2019multicast} is employed whereby the first stage precoder is designed based on the feeder link channel while the second stage precoder is designed based on the user link channel. The key benefit of this two-stage precoding design is that the feeder link interference and the user link interference can be mitigated independently such that a cleaner signal that is free from feeder link interference can be used during the second stage precoder design to mitigate the multibeam interference at the user link more effectively.

In order to carry out this two-stage precoding, the precoder for each (common or private) symbol stream is split into two parts, i.e. $\mathbf{p}_{c,l} = \mathbf{W}_l \mathbf{v}_{c,l}$ and $\mathbf{p}_m = \mathbf{W}_{\lambda \left( m \right)} \mathbf{v}_m$. During the first stage of the precoder design problem, the first stage precoders $\mathbf{W} = \left[ \mathbf{W}_1, \ldots, \mathbf{W}_L \right]$ and OBP matrices $\mathbf{R} = \left[ \mathbf{R}_1, \ldots, \mathbf{R}_L \right]$ are jointly optimised to tackle the feeder link interference. Once the optimal solutions are obtained, the effective feeder link channel between gateway $l$ and feed cluster $i$ is $\overline{\mathbf{F}}_{i,l} = \mathbf{R}_i \mathbf{F}_{i,l} \mathbf{W}_l$. The new effective channel between gateway $j$ and user $k$ is now $\overline{\mathbf{h}}^H_{j,k} = \sum_{l=1}^L \mathbf{h}^H_{l,k} \overline{\mathbf{F}}_{l,j} = \sum_{l=1}^L \mathbf{h}^H_{l,k} \mathbf{R}_l \mathbf{F}_{l,j} \mathbf{W}_j$.

After introducing the precoders for the two stages, the received signal at user $k$ can be written as
\begin{subequations}
\begin{align}
    y_k & = \sum_{l=1}^L \sum_{j=1}^L \mathbf{h}^H_{l,k} \mathbf{R}_l \mathbf{F}_{l,j} \mathbf{W}_j \mathbf{v}_{c,j}s_{c,j} + \sum_{l=1}^L \sum_{j=1}^L \sum_{m \in \mathcal{B}_j} \mathbf{h}^H_{l,k} \mathbf{R}_l \mathbf{F}_{l,j} \mathbf{W}_j \mathbf{v}_m s_m + \sum_{l=1}^L \mathbf{h}^H_{l,k} \mathbf{R}_l \mathbf{n}_l + n_k \\
    & = \sum_{j=1}^L \overline{\mathbf{h}}^H_{j,k} \mathbf{v}_{c,j}s_{c,j} + \sum_{j=1}^L \sum_{m \in \mathcal{B}_j} \overline{\mathbf{h}}^H_{j,k} \mathbf{v}_m s_m + \overline{n}_k. \label{equ_received_signal_obp}
\end{align}
\end{subequations}

To obtain the group-rate for each group, the formulae used to calculate the SINRs of decoding the desired common and private stream can be reapplied but with the second stage precoders $\mathbf{V} = \left[ \mathbf{v}_{c,1}, \ldots, \mathbf{v}_{c,L}, \mathbf{v}_1, \ldots, \mathbf{v}_M \right]$ used instead since (\ref{equ_received_signal_obp}) is similar to (\ref{equ_received_signal}). 

Similar to the case with no OBP in the previous subsection, the deterministic approximation of the MMF problem for the second stage precoder design using the RS strategy at multiple gateways under imperfect CSIT with per-antenna power constraints is formulated as

$\mathcal{F}^{ \left( S \right)}_{RS_{OBP}} \left( P \right)$:
\begin{subequations}
\begin{align}
\max_{\overline{\mathbf{c}},\mathbf{V}} \quad & \min_{m \in \mathcal{M}} \left(\overline{C}_m + \overline{r}_m\right) \\
s.t. \quad & \overline{R}_{c,k}^{ \left( S \right)} \geq \sum_{i \in \mathcal{B}_l} \overline{C}_i,\ \forall k \in \mathcal{G}_m, m \in \mathcal{B}_l, l \in \mathcal{L}, \\
& \overline{C}_m \geq 0,\ \forall m \in \mathcal{M}, \\
& \sum_{j=1}^L \mathbf{v}^H_{c,j} \widehat{\overline{\mathbf{F}}}^H_{l,j} \mathbf{D}_{f \left( n,\mathcal{B}_l \right)} \widehat{\overline{\mathbf{F}}}_{l,j} \mathbf{v}_{c,j} + \sum_{m \in \mathcal{M}} \mathbf{v}^H_m \widehat{\overline{\mathbf{F}}}^H_{l,\lambda \left( m \right)} \mathbf{D}_{f \left( n,\mathcal{B}_l \right)} \widehat{\overline{\mathbf{F}}}_{l,\lambda \left( m \right)} \mathbf{v}_m \nonumber \\
& + \sigma^2_n \left[ \mathbf{R}_l \mathbf{R}^H_l \right]_{f \left( n,\mathcal{B}_l \right)f \left( n,\mathcal{B}_l \right)} \leq P_n,\ \forall n \in \mathcal{B}_l, l \in \mathcal{L}.
\end{align}
\end{subequations}

Once again, the average common rate portions and the second stage precoders are jointly optimised by solving this problem. In contrast to $\mathcal{F}^{\left( S \right)}_{RS_{feeder}} \left( P \right)$, the realisation set for the effective channel between the gateways and the users $\overline{\mathbb{H}}^{ \left( S \right)}$ is formed using the optimised effective feeder link channel estimate $\widehat{\overline{\mathbf{F}}}$ obtained after the first stage precoding and the user link channel realisation set $\mathbb{H}^{ \left( S \right)}$. Also, the optimised first stage precoders and OBP receiving filter matrices are used in the satellite transmit power constraint. The optimisation of the first and second stage precoding will be illustrated in the following section.

\section{Optimisation}
The optimisation problems formulated thus far are very challenging to be solved in their current form because they contain non-convex coupled sum-rate expressions \cite{joudeh2017rate}. The WMMSE approach introduced in \cite{christensen2009weighted} is extended in this paper such that the optimisation problems can be reformulated into equivalent augmented WMSE problems which are convex and solvable. 

In this section, we first introduce a WMMSE approach together with AO algorithm to solve the formulated MMF problem $\mathcal{F}^{\left( S \right)}_{RS_{feeder}} \left( P \right)$ in the presence of noisy feeder link with interference. Then, with the use of OBP, the first stage precoding and OBP receiving filter matrices are optimised together using a MMSE metric. The second stage precoding is optimised by solving the formulated MMF problem $\mathcal{F}^{ \left( S \right)}_{RS_{OBP}} \left( P \right)$ based on the same WMMSE and AO algorithm.

\subsection{WMMSE Approach}
The stream that needs to be decoded first at user $k$ is the desired common stream $s_{c,\lambda \left[ \mu \left( k \right) \right]}$. The estimate of the common stream is $\widehat{s}_{c,\lambda \left[ \mu \left( k \right) \right]}$ and it is obtained using a scalar equaliser, $g_{c,k}$ such that $\widehat{s}_{c,\lambda \left[ \mu \left( k \right) \right]} = g_{c,k} y_k$ \cite{joudeh2017rate}. After decoding and removing the desired common stream from the received signal, the estimate of the desired private stream $\widehat{s}_{\mu \left( k \right)}$ is also obtained using a scalar equaliser, $g_k$ such that $\widehat{s}_{\mu \left( k \right)} = g_k \left( y_k - \overline{\mathbf{h}}^H_{\lambda \left[ \mu \left( k \right) \right],k} \mathbf{p}_{c,\lambda \left[ \mu \left( k \right) \right]} s_{c,\lambda \left[ \mu \left( k \right) \right]} \right)$ \cite{joudeh2017rate}.

Based on the user's received signal, the average received power is $T_{c,k} = \left|\overline{\mathbf{h}}^H_{\lambda \left[ \mu \left( k \right) \right],k} \mathbf{p}_{c,\lambda \left[ \mu \left( k \right) \right]}\right|^2 + \left|\overline{\mathbf{h}}^H_{\lambda \left[ \mu \left( k \right) \right],k} \mathbf{p}_{\mu \left( k \right)}\right|^2 + \sum_{m \in \mathcal{B}_{\lambda \left[ \mu \left( k \right) \right]} \setminus \mu \left( k \right)} \left|\overline{\mathbf{h}}^H_{\lambda \left[ \mu \left( k \right) \right],k} \mathbf{p}_m\right|^2 \nonumber + \sum_{j \in \mathcal{L} \setminus \lambda \left[ \mu \left( k \right) \right]} \left[\left|\overline{\mathbf{h}}^H_{j,k} \mathbf{p}_{c,j}\right|^2 + \sum_{m \in \mathcal{B}_j}\left|\overline{\mathbf{h}}^H_{j,k} \mathbf{p}_m\right|^2\right] + \overline{\sigma}^2_{n,k}$.

After SIC is carried out, the observed power is defined as $T_k = T_{c,k} - \left|\overline{\mathbf{h}}^H_{\lambda \left[ \mu \left( k \right) \right],k} \mathbf{p}_{c,\lambda \left[ \mu \left( k \right) \right]}\right|^2$. The portion of the powers contributed by the interference in $T_{c,k}$ and $T_k$ can also be identified and they are $I_{c,k} = T_k$ and $I_k = T_k - \left|\overline{\mathbf{h}}^H_{\lambda \left[ \mu \left( k \right) \right],k} \mathbf{p}_{\mu \left( k \right)}\right|^2$ respectively.

According to \cite{joudeh2017rate}, the stream estimates and the original streams are compared. The MSE for the desired common stream $\epsilon_{c,k} = \mathbb{E} \left\{\left|\widehat{s}_{c,\lambda \left[ \mu \left( k \right) \right]} - s_{c,\lambda \left[ \mu \left( k \right) \right]}\right|^2\right\}$ and the MSE for the desired private stream $\epsilon_k = \mathbb{E} \left\{\left|\widehat{s}_{\mu \left( k \right)} - s_{\mu \left( k \right)}\right|^2\right\}$ can be expanded and written as
\begin{subequations}
\begin{gather}
    \epsilon_{c,k} = \left|g_{c,k}\right|^2 T_{c,k} - 2\mathcal{R}\left\{g_{c,k} \overline{\mathbf{h}}^H_{\lambda \left[ \mu \left( k \right) \right],k} \mathbf{p}_{c,\lambda \left[ \mu \left( k \right) \right]}\right\} + 1, \\
    \epsilon_k = \left|g_k\right|^2 T_k - 2\mathcal{R}\left\{g_k \overline{\mathbf{h}}^H_{\lambda \left[ \mu \left( k \right) \right],k} \mathbf{p}_{\mu \left( k \right)}\right\} + 1.
\end{gather}
\end{subequations}

In order to minimise the MSE, the first derivative of the MSE with respect to the equaliser is set to 0, i.e. $\frac{\partial \epsilon_{c,k}}{\partial g_{c,k}} = 0$ and $\frac{\partial \epsilon_k}{\partial g_k} = 0$ \cite{mao2018rate}. By solving the two equations, the optimal equalisers that would result in the MMSEs are $g_{sc,k}^{MMSE} = \mathbf{p}_{c,\lambda \left[ \mu \left( k \right) \right]}^H \overline{\mathbf{h}}_{\lambda \left[ \mu \left( k \right) \right],k} T_{c,k}^{-1}$ and $g_k^{MMSE} = \mathbf{p}_{\mu \left( k \right)}^H \overline{\mathbf{h}}_{\lambda \left[ \mu \left( k \right) \right],k} T_k^{-1}$. After obtaining the optimal equalisers, they are substituted back into the equations for the MSEs such that the MMSEs are written as
\begin{subequations}
\begin{gather}
    \epsilon_{c,k}^{MMSE} = \min_{g_{c,k}} \epsilon_{c,k} = T_{c,k}^{-1} I_{c,k}, \\
    \epsilon_k^{MMSE} = \min_{g_k} \epsilon_k = T_k^{-1} I_k.
\end{gather}
\end{subequations}

Based on the MMSE obtained for the desired common stream and the desired private stream, it can be seen that the SINR for the two streams can be represented in terms of their respective MMSE, i.e. $\gamma_{c,k} = \frac{1}{\epsilon_{c,k}^{MMSE}} - 1$ and $\gamma_k = \frac{1}{\epsilon_k^{MMSE}} - 1$ \cite{joudeh2017rate}. Since $R_{c,k} = \log_2 \left( 1+\gamma_{c,k} \right)$ and $R_k = \log_2 \left( 1+\gamma_k \right)$, it can be seen that the achievable rates can also be represented in terms of the MMSEs, i.e. $R_{c,k} = - \log_2\left(\epsilon_{c,k}^{MMSE}\right)$ and $R_k = - \log_2\left(\epsilon_k^{MMSE}\right)$ \cite{joudeh2017rate}.

The augmented WMSE for the desired common stream and the desired private stream are the key foundation to the solution and they are defined as
\begin{subequations}
\begin{gather}
    \xi_{c,k} = u_{c,k}\epsilon_{c,k} - \log_2 \left( u_{c,k} \right), \\
    \xi_k = u_k\epsilon_k - \log_2 \left( u_k \right),
\end{gather}
\end{subequations}
where $u_{c,k}$ and $u_k$ are the weights associated with the respective MSE and they are bigger than zero \cite{joudeh2017rate}.

To obtain the optimal equalisers for the augmented WMSEs, the equations $\frac{\partial \xi_{c,k}}{\partial g_{c,k}} = 0$ and $\frac{\partial \xi_k}{\partial g_k} = 0$ are solved and the optimal equalisers are found to be the same as the MMSE equalisers \cite{mao2018rate}. Substituting the optimal equalisers into the equations for the augmented WMSEs, the equations are now written as
\begin{subequations}
\begin{gather}
    \xi_{c,k}\left(g_{c,k}^{MMSE}\right) = \min_{g_{c,k}} \xi_{c,k} = u_{c,k}\epsilon_{c,k}^{MMSE} - \log_2 \left( u_{c,k} \right), \\
    \xi_k\left(g_k^{MMSE}\right) = \min_{g_k} \xi_k = u_k\epsilon_k^{MMSE} - \log_2 \left( u_k \right).
\end{gather}
\end{subequations}

To obtain the optimal weights for the augmented WMSEs, the equations $\frac{\partial \xi_{c,k}\left(g_{c,k}^{MMSE}\right)}{\partial u_{c,k}} = 0$ and $\frac{\partial \xi_k\left(g_k^{MMSE}\right)}{\partial u_k} = 0$ are solved and the optimal weights are $u_{c,k}^{MMSE} = \left(\epsilon_{c,k}^{MMSE}\right)^{-1}$ and $u_k^{MMSE} = \left(\epsilon_k^{MMSE}\right)^{-1}$ \cite{mao2018rate}.

According to \cite{joudeh2017rate}, substituting the optimal weights back into the equations for the augmented WMSEs with optimal equalisers leads to a Rate-WMMSE relationship and this can be written as
\begin{subequations}
\begin{gather}
    \xi_{sc,k}^{MMSE} = \min_{u_{sc,k},g_{sc,k}} \xi_{sc,k} = 1 - R_{sc,k}, \\
    \xi_k^{MMSE} = \min_{u_k,g_k} \xi_k = 1 - R_k.
\end{gather}
\end{subequations}

An average version of the Rate-WMMSE relationship can be obtained by taking the expectation over the conditional distribution of $\overline{\mathbf{H}}$ given $\widehat{\overline{\mathbf{H}}}$ \cite{joudeh2016sum}. The relationship is now expressed as
\begin{subequations}
\begin{gather}
    \overline{\xi}_{c,k}^{MMSE} = \mathbb{E}_{\overline{\mathbf{H}} \mid \widehat{\overline{\mathbf{H}}}} \left\{\min_{u_{c,k},g_{c,k}} \overline{\xi}_{c,k} \mid \widehat{\overline{\mathbf{H}}} \right\} = 1 - \overline{R}_{c,k}, \\
    \overline{\xi}_k^{MMSE} = \mathbb{E}_{\overline{\mathbf{H}} \mid \widehat{\overline{\mathbf{H}}}} \left\{\min_{u_k,g_k} \overline{\xi}_k \mid \widehat{\overline{\mathbf{H}}} \right\} = 1 - \overline{R}_k.
\end{gather}
\end{subequations}

In order to transform the stochastic problem into a deterministic problem, SAA is used once again to formulate the augmented WMSE, equaliser and weight associated with the $s$-th realisation in $\overline{\mathbb{H}}^{ \left( S \right)}$. The set of equalisers and weights for the desired common stream and the desired private stream for a user are denoted as $\mathbf{g}_{c,k} = \left[g_{c,k}^{\left( 1 \right)},\ldots,g_{c,k}^{ \left( S \right)}\right]^T$, $\mathbf{g}_k = \left[g_k^{\left( 1 \right)},\ldots,g_k^{ \left( S \right)}\right]^T$, $\mathbf{u}_{c,k} = \left[u_{c,k}^{\left( 1 \right)},\ldots,u_{c,k}^{ \left( S \right)}\right]^T$ and $\mathbf{u}_k = \left[u_k^{\left( 1 \right)},\ldots,u_k^{ \left( S \right)}\right]^T$.

For compactness, $\mathbf{G} = \left\{ \mathbf{g}_{c,k}, \mathbf{g}_k \mid k \in \mathcal{K} \right\}$ contains the equaliser vector from all users while $\mathbf{U} = \left\{ \mathbf{u}_{c,k}, \mathbf{u}_k \mid k \in \mathcal{K} \right\}$ contains the weight vector from all users. According to \cite{joudeh2016sum}, the average augmented WMSEs are then approximated using their sample average functions (SAFs) such that
\begin{equation}
    \overline{\xi}_{c,k}^{ \left( S \right)} = \frac{1}{S} \sum_{s=1}^S \xi_{c,k}^{ \left( s \right)} \quad \mathrm{and} \quad \overline{\xi}_k^{ \left( S \right)} = \frac{1}{S} \sum_{s=1}^S \xi_k^{ \left( s \right)}.
\end{equation}

Also, when the optimal equalisers $\mathbf{G}^{MMSE}$ and weights $\mathbf{U}^{MMSE}$ are used in the approximated average augmented WMSEs, the deterministic SAA version of the Rate-WMMSE relationship can be written as
\begin{subequations}
\begin{gather}
    \overline{\xi}_{c,k}^{MMSE \left( S \right)} = \min_{\mathbf{u}_{c,k},\mathbf{g}_{c,k}} \overline{\xi}_{c,k}^{ \left( S \right)} = 1 - \overline{R}_{c,k}^{ \left( S \right)}, \\
    \overline{\xi}_k^{MMSE \left( S \right)} = \min_{\mathbf{u}_k,\mathbf{g}_k} \overline{\xi}_k^{ \left( S \right)} = 1 - \overline{R}_k^{ \left( S \right)}.
\end{gather}
\end{subequations}

$\mathcal{F}_{RS_{feeder}}^{ \left( S \right)} \left( P \right)$ can now be reformulated into the equivalent WMSE problem and written as

$\mathcal{W}_{RS_{feeder}}^{ \left( S \right)} \left( P \right)$:
\begin{subequations}
\begin{align}
\max_{\overline{r}_g, \overline{\mathbf{r}}, \overline{\mathbf{c}},\mathbf{P},\mathbf{G},\mathbf{U}} \quad & \overline{r}_g \\
s.t. \quad & \overline{C}_m + \overline{r}_m \geq \overline{r}_g,\ \forall m \in \mathcal{M}, \\
& 1 - \overline{\xi}_i^{ \left( S \right)} \geq \overline{r}_m,\ \forall i \in \mathcal{G}_m, m \in \mathcal{M}, \\
& 1 - \overline{\xi}_{c,k}^{ \left( S \right)} \geq \sum_{i \in \mathcal{B}_l} \overline{C}_i,\ \forall k \in \mathcal{G}_m, m \in \mathcal{B}_l, l \in \mathcal{L}, \\
& \overline{C}_m \geq 0,\ \forall m \in \mathcal{M}, \\
& \sum_{j=1}^L \mathbf{p}^H_{c,j} \widehat{\mathbf{F}}^H_{l,j} \mathbf{D}_{f \left( n,\mathcal{B}_l \right)} \widehat{\mathbf{F}}_{l,j} \mathbf{p}_{c,j} + \sum_{m \in \mathcal{M}} \mathbf{p}^H_m \widehat{\mathbf{F}}^H_{l,\lambda \left( m \right)} \mathbf{D}_{f \left( n,\mathcal{B}_l \right)} \widehat{\mathbf{F}}_{l,\lambda \left( m \right)} \mathbf{p}_m \nonumber \\
& + \sigma^2_n \leq P_n,\ \forall n \in \mathcal{B}_l, l \in \mathcal{L}.
\end{align}
\end{subequations}
where $\overline{r}_g$ and $\overline{\mathbf{r}} = \left[ \overline{r}_1, \ldots, \overline{r}_M \right]$ are auxiliary variables.

The optimisation problem is still non-convex in the joint set of optimisation variables (equalisers, weights, precoders) but the problem is convex in each variable while the other two variables are fixed therefore an AO algorithm can be used to solve the problem \cite{joudeh2016sum}. There are two steps involved in the AO algorithm and they are described in full detail next.

\subsubsection{Updating $\mathbf{G}$ and $\mathbf{U}$ for a given $\mathbf{P}$}
The equalisers and the weights are updated with the precoders obtained from the previous iteration by using the closed form expressions obtained, i.e. $\mathbf{G} = \mathbf{G}^{MMSE}\left(\mathbf{P}^{\left[ n-1 \right]}\right)$ and $\mathbf{U} = \mathbf{U}^{MMSE}\left(\mathbf{P}^{\left[ n-1 \right]}\right)$ \cite{yin2020rate}. According to \cite{joudeh2016sum}, additional terms are introduced to facilitate the formation of the precoder optimisation problem before moving on to the next step in the algorithm and they are
\begin{align}
    t_{c,k}^{ \left( s \right)} & = u_{c,k}^{ \left( s \right)}\left|g_{c,k}^{ \left( s \right)}\right|^2 && \mathrm{and} & t_k^{ \left( s \right)} & = u_k^{ \left( s \right)}\left|g_k^{ \left( s \right)}\right|^2, \\
    \Psi_{c,j,k}^{ \left( s \right)} & = t_{c,k}^{ \left( s \right)} \overline{\mathbf{h}}_{j,k}^{ \left( s \right)} \overline{\mathbf{h}}_{j,k}^{ \left( s \right)H} && \mathrm{and} & \Psi_{j,k}^{ \left( s \right)} & = t_k^{ \left( s \right)} \overline{\mathbf{h}}_{j,k}^{ \left( s \right)} \overline{\mathbf{h}}_{j,k}^{ \left( s \right)H}, \\
    \mathbf{f}_{c,k}^{ \left( s \right)} & = u_{c,k}^{ \left( s \right)} \overline{\mathbf{h}}_{\lambda \left[ \mu \left( k \right) \right],k}^{ \left( s \right)} g_{c,k}^{ \left( s \right)H} && \mathrm{and} & \mathbf{f}_k^{ \left( s \right)} & = u_k^{ \left( s \right)} \overline{\mathbf{h}}_{\lambda \left[ \mu \left( k \right) \right],k}^{ \left( s \right)} g_k^{ \left( s \right)H}, \\
    v_{c,k}^{ \left( s \right)} & = \log_2 \left(u_{c,k}^{ \left( s \right)}\right) && \mathrm{and} & v_k^{ \left( s \right)} & = \log_2 \left(u_k^{ \left( s \right)}\right).
\end{align}

The corresponding SAFs together for the above terms and the weights are used to form $\overline{t}_{c,k}^{ \left( S \right)}$, $\overline{t}_k^{ \left( S \right)}$, $\overline{\Psi}_{c,j,k}^{ \left( S \right)}$, $\overline{\Psi}_{j,k}^{ \left( S \right)}$, $\overline{\mathbf{f}}_{c,k}^{ \left( S \right)}$, $\overline{\mathbf{f}}_k^{ \left( S \right)}$, $\overline{v}_{c,k}^{ \left( S \right)}$, $\overline{v}_k^{ \left( S \right)}$, $\overline{u}_{c,k}^{ \left( S \right)}$, $\overline{u}_k^{ \left( S \right)}$ by averaging the terms over all $S$ realisations \cite{joudeh2016sum}. 

\subsubsection{Updating $\mathbf{P}$ and auxiliary variables for a given $\mathbf{G}$ and $\mathbf{U}$}
The precoders and the auxiliary variables are updated with the newly updated equalisers and weights from the previous step. By substituting the newly updated $\mathbf{G}$ and $\mathbf{U}$ into $\mathcal{W}_{RS_{feeder}}^{ \left( S \right)} \left( P \right)$, the problem of updating $\mathbf{P}$ turns into

$\mathcal{W}_{RS_{feeder}}^{ \left[ n \right]} \left( P \right)$:
\begin{subequations}
\begin{align}
\max_{\overline{r}_g, \overline{\mathbf{r}}, \overline{\mathbf{c}},\mathbf{P}} \quad & \overline{r}_g \\
s.t. \quad & \overline{C}_m + \overline{r}_m \geq \overline{r}_g,\ \forall m \in \mathcal{M}, \\
& 1 - \overline{r}_m \geq \sum_{j \in \mathcal{L} \setminus \lambda \left[ \mu \left( i \right) \right]} \mathbf{p}_{c,j}^H \overline{\Psi}_{j,i}^{ \left( S \right)} \mathbf{p}_{c,j} + \sum_{m \in \mathcal{M}} \mathbf{p}_m^H \overline{\Psi}_{\lambda \left( m \right),i}^{ \left( S \right)} \mathbf{p}_m + \widehat{\overline{\sigma}}_{n,i}^2 \overline{t}_i^{ \left( S \right)} \nonumber \\
& - 2\mathcal{R}\left(\overline{\mathbf{f}}_i^{ \left( S \right)H} \mathbf{p}_{\mu \left( i \right)}\right) + \overline{u}_i^{ \left( S \right)} - \overline{v}_i^{ \left( S \right)},\ \forall i \in \mathcal{G}_m, m \in \mathcal{M}, \\
& 1 - \sum_{i \in \mathcal{B}_l} \overline{C}_i \geq \sum_{j \in \mathcal{L}} \mathbf{p}_{c,j}^H \overline{\Psi}_{c,j,k}^{ \left( S \right)} \mathbf{p}_{c,j} + \sum_{m \in \mathcal{M}} \mathbf{p}_m^H \overline{\Psi}_{c,\lambda \left( m \right),k}^{ \left( S \right)} \mathbf{p}_m + \widehat{\overline{\sigma}}_{n,k}^2 \overline{t}_{c,k}^{ \left( S \right)} \nonumber \\
& - 2\mathcal{R}\left(\overline{\mathbf{f}}_{c,k}^{ \left( S \right)H} \mathbf{p}_{c,\lambda \left[ \mu \left( k \right) \right]}\right) + \overline{u}_{c,k}^{ \left( S \right)} - \overline{v}_{c,k}^{ \left( S \right)},\ \forall k \in \mathcal{G}_m, m \in \mathcal{B}_l, l \in \mathcal{L}, \\
& \overline{C}_m \geq 0,\ \forall m \in \mathcal{M}, \\
& \sum_{j=1}^L \mathbf{p}^H_{c,j} \widehat{\mathbf{F}}^H_{l,j} \mathbf{D}_{f \left( n,\mathcal{B}_l \right)} \widehat{\mathbf{F}}_{l,j} \mathbf{p}_{c,j} + \sum_{m \in \mathcal{M}} \mathbf{p}^H_m \widehat{\mathbf{F}}^H_{l,\lambda \left( m \right)} \mathbf{D}_{f \left( n,\mathcal{B}_l \right)} \widehat{\mathbf{F}}_{l,\lambda \left( m \right)} \mathbf{p}_m \nonumber \\
& + \sigma^2_n \leq P_n,\ \forall n \in \mathcal{B}_l, l \in \mathcal{L}.
\end{align}
\end{subequations}

This is a convex quadratically constrained quadratic problem (QCQP) \cite{joudeh2016sum} and it can be solved using interior-point methods \cite{boyd2004convex}. After obtaining a new set of precoders, the first step of the AO algorithm is carried out again. The optimisation variables are updated alternately and this whole process of repeating the two steps will be carried out until the MMF rate obtained from the second step converges.

\begin{algorithm}[t]
\caption{Alternating Optimisation}
\label{algo_1}
\begin{algorithmic}[1]
\State \textbf{Initialise}: $n \leftarrow 0, \mathbf{P}, \mathcal{W}_{RS_{feeder}}^{ \left[ n \right]} \left( P \right) \leftarrow 0$
\State \textbf{repeat}
\State $\quad n \leftarrow n+1, \mathbf{P}^{\left[ n-1 \right]} \leftarrow \mathbf{P}$
\State $\quad \mathbf{G} \leftarrow \mathbf{G}^{MMSE}\left(\mathbf{P}^{\left[ n-1 \right]}\right), \mathbf{U} \leftarrow \mathbf{U}^{MMSE}\left(\mathbf{P}^{\left[ n-1 \right]}\right)$
\State $\quad \text{update}\ \overline{t}_{c,k}^{ \left( S \right)}, \overline{t}_k^{ \left( S \right)}, \overline{\Psi}_{c,j,k}^{ \left( S \right)}, \overline{\Psi}_{j,k}^{ \left( S \right)}, \overline{\mathbf{f}}_{c,k}^{ \left( S \right)}, \overline{\mathbf{f}}_k^{ \left( S \right)}, \overline{v}_{c,k}^{ \left( S \right)}, \overline{v}_k^{ \left( S \right)},$
\Statex $\quad \overline{u}_{c,k}^{ \left( S \right)}, \overline{u}_k^{ \left( S \right)},\ \text{for all}\ k \in \mathcal{K}$
\State $\quad \mathbf{P} \leftarrow \arg \mathcal{W}_{RS_{feeder}}^{ \left[ n \right]} \left( P \right)$ using updated $\mathbf{G}$ and $\mathbf{U}$
\State \textbf{until} $\left|\mathcal{W}_{RS_{feeder}}^{ \left[ n \right]} \left( P \right) - \mathcal{W}_{RS_{feeder}}^{ \left[ n-1 \right]} \left( P \right)\right| \leq \epsilon$
\end{algorithmic}
\end{algorithm}

Full details of the AO algorithm implemented is described in Algorithm \ref{algo_1}. Note that $\epsilon$ is the tolerance level for convergence. The precoder for a stream is initialised by finding the left singular vector that corresponds to the biggest singular value for the composite channel made from the channels of the intended users. The singular vector is then multiplied with the amount of power obtained after equally sharing the total transmit power available with the other precoders.

\subsection{Two-Stage Precoding Design}
\subsubsection{First Stage}
Taking a closer look at the feeder link, an equivalent MIMO interference channel \cite{wang2019multicast} can be obtained by treating the multiple gateways as transmitters with multiple antennas and the satellite receivers as users with multiple antennas. With this concept in mind and under imperfect CSIT, the transmitted signals and the estimated stream vector from the received signals can be rewritten as
\begin{align}
    \mathbf{x}_l & = \mathbf{P}_l\mathbf{s}_l = \mathbf{W}_l \mathbf{V}_l \mathbf{s}_l = \mathbf{W}_l\mathbf{d}_l,\ \forall l \in \mathcal{L}, \\
    \widehat{\mathbf{d}}_l & = \mathbf{R}_l \mathbf{r}_l = \mathbf{R}_l \sum_{i=1}^L \mathbf{F}_{l,i} \mathbf{W}_i \mathbf{d}_i + \mathbf{R}_l \mathbf{n}_l \nonumber \\
    & = \mathbf{R}_l \sum_{i=1}^L \left( \widehat{\mathbf{F}}_{l,i} + \widetilde{\mathbf{F}}_{l,i} \right) \mathbf{W}_i \mathbf{d}_i + \mathbf{R}_l \mathbf{n}_l,\ \forall l \in \mathcal{L},
\end{align}
where $\mathbf{W}_l$ is the precoding matrix at the transmitter, $\mathbf{d}_l$ is the data stream vector, $\mathbf{R}_l$ is the receiving filter matrix and $\widehat{\mathbf{d}}_l$ is the estimated data stream vector.

According to \cite{shen2010mse}, the estimated stream vector and the original stream vector at each receiver are compared to obtain the MSE and it can be written as
\begin{subequations}
\begin{align}
    \mathrm{MSE}_l & = \mathbb{E} \left\{\left\| \widehat{\mathbf{d}}_l - \mathbf{d}_l \right\|^2\right\} \\
    & = \mathrm{Tr} \Bigg[ \mathbf{R}_l \left( \sum_{i=1}^L \widehat{\mathbf{F}}_{l,i} \mathbf{W}_i \mathbf{W}^H_i \widehat{\mathbf{F}}^H_{l,i} \right) \mathbf{R}^H_l + \sigma_e^2 \mathrm{Tr} \left( \mathbf{R}_l \mathbf{R}^H_l \right)\sum_{i=1}^L \mathbf{W}_i \mathbf{W}^H_i \nonumber \\
    & - \mathbf{R}_l \widehat{\mathbf{F}}_{l,l} \mathbf{W}_l - \mathbf{W}^H_l \widehat{\mathbf{F}}^H_{l,l} \mathbf{R}^H_l + \mathbf{I} + \sigma_n^2 \mathbf{R}_l \mathbf{R}^H_l \Bigg].
\end{align}
\end{subequations}

The sum MSE is then defined as $\mathrm{MSE}_{\mathrm{sum}} = \sum_{l=1}^L \mathrm{MSE}_l$. In order to minimise the sum MSE, the partial derivative of the sum MSE with respect to the matrix $\mathbf{W}_l$ and the matrix $\mathbf{R}_l$ is set to 0, i.e. $\frac{\partial \mathrm{MSE}_{\mathrm{sum}}}{\partial \mathbf{W}_l} = 0$ and $\frac{\partial \mathrm{MSE}_{\mathrm{sum}}}{\partial \mathbf{R}_l} = 0$ \cite{shen2010mse}. By solving the two equations, the optimal first stage precoders and the optimal OBP matrices are
\begin{align}
    \mathbf{R}^{MMSE}_l & = \mathbf{W}^H_l \widehat{\mathbf{F}}^H_{l,l} \left( \sum_{i=1}^L \widehat{\mathbf{F}}_{l,i} \mathbf{W}_i \mathbf{W}^H_i \widehat{\mathbf{F}}^H_{l,i} + \sigma_n^2 \mathbf{I} + \sigma_e^2 \mathbf{I} \right)^{-1}, \\
    \mathbf{W}^{MMSE}_l & = \left[ \sum_{i=1}^L \widehat{\mathbf{F}}^H_{i,l} \mathbf{R}^H_i \mathbf{R}_i \widehat{\mathbf{F}}_{i,l} + \sigma_e^2 \mathrm{Tr} \left(\sum_{i=1}^L \mathbf{R}^H_i \mathbf{R}_i \right) \mathbf{I} \right]^{-1} \widehat{\mathbf{F}}^H_{l,l} \mathbf{R}^H_l.
\end{align}

For compactness, $\mathbf{R} = \left\{ \mathbf{R}_l \mid l \in \mathcal{L} \right\}$ contains all the OBP receiving filter matrices while $\mathbf{W} = \left\{ \mathbf{W}_l \mid l \in \mathcal{L} \right\}$ contains the first stage precoder from all gateways.

The AO algorithm is used to solve this problem because the optimisation problem is once again convex for each variable while keeping the other constant but non-convex for the joint set \cite{shen2010mse}. During each iteration, the set of OBP receiving filter matrices are first updated with a given set of first stage percoders. Then, the set of first stage precoders are updated with a given set of OBP receiving filter matrices. These two steps will then be repeated until the sum MSE converges.

\begin{algorithm}[t]
\caption{First Stage Precoder Design with OBP}
\label{algo_2}
\begin{algorithmic}[1]
\State \textbf{Initialise}: $n \leftarrow 0, \mathbf{W}^{\left[ n \right]}, \mathrm{MSE}^{\left[ n \right]}_{\mathrm{sum}} \leftarrow 0$
\State \textbf{repeat}
\State $\quad n \leftarrow n+1$
\State $\quad \mathbf{R}^{\left[ n \right]} \leftarrow \mathbf{R}^{MMSE}\left(\mathbf{W}^{\left[ n-1 \right]}\right)$
\State $\quad \mathbf{W}^{\left[ n \right]} \leftarrow \mathbf{W}^{MMSE}\left(\mathbf{R}^{\left[ n \right]}\right)$
\State $\quad \text{update}\ \mathrm{MSE}_l,\ \text{for all}\ l \in \mathcal{L}$
\State \textbf{until} $\left|\mathrm{MSE}^{\left[ n \right]}_{\mathrm{sum}} - \mathrm{MSE}^{\left[ n-1 \right]}_{\mathrm{sum}}\right| \leq \epsilon$
\end{algorithmic}
\end{algorithm}

Full details of the algorithm implemented for the first stage precoder design is described in Algorithm \ref{algo_2}. The first stage precoder $\mathbf{W}_l$ for gateway $l$ is initialised with its entries randomly drawn from a complex Gaussian distribution with zero mean and unit variance \cite{shen2010mse}. Note that when perfect CSIT is available at the gateways, Algorithm \ref{algo_2} and the closed form expressions for $\mathbf{R}^{MMSE}$ and $\mathbf{W}^{MMSE}$ can be reused during the first stage precoder design by just setting $\sigma_e^2$ to $0$ and using the actual feeder link channels in place of the estimates.

\subsubsection{Second Stage}
After the first stage, the optimised first stage precoders and OBP receiving filter matrices are used to formulate the effective feeder link channel and thereafter the effective channel from a gateway to a user as described in Section \ref{sec_obp}. As for the second stage, the deterministic MMF problem $\mathcal{F}_{RS_{OBP}}^{ \left( S \right)} \left( P \right)$ is still non convex and therefore the WMMSE approach described earlier is used again to convert it into a solvable convex problem, which can be written as

$\mathcal{W}_{RS_{OBP}}^{ \left( S \right)} \left( P \right)$:
\begin{subequations}
\begin{align}
\max_{\overline{r}_g, \overline{\mathbf{r}}, \overline{\mathbf{c}}, \mathbf{V},\mathbf{G},\mathbf{U}} \quad & \overline{r}_g \\
s.t. \quad & \overline{C}_m + \overline{r}_m \geq \overline{r}_g,\ \forall m \in \mathcal{M}, \\
& 1 - \overline{\xi}_i^{ \left( S \right)} \geq \overline{r}_m,\ \forall i \in \mathcal{G}_m, m \in \mathcal{M}, \\
& 1 - \overline{\xi}_{c,k}^{ \left( S \right)} \geq \sum_{i \in \mathcal{B}_l} \overline{C}_i,\ \forall k \in \mathcal{G}_m, m \in \mathcal{B}_l, l \in \mathcal{L}, \\
& \overline{C}_m \geq 0,\ \forall m \in \mathcal{M}, \\
& \sum_{j=1}^L \mathbf{v}^H_{c,j} \widehat{\overline{\mathbf{F}}}^H_{l,j} \mathbf{D}_{f \left( n,\mathcal{B}_l \right)} \widehat{\overline{\mathbf{F}}}_{l,j} \mathbf{v}_{c,j} + \sum_{m \in \mathcal{M}} \mathbf{v}^H_m \widehat{\overline{\mathbf{F}}}^H_{l,\lambda \left( m \right)} \mathbf{D}_{f \left( n,\mathcal{B}_l \right)} \widehat{\overline{\mathbf{F}}}_{l,\lambda \left( m \right)} \mathbf{v}_m \nonumber \\
& + \sigma^2_n \left[ \mathbf{R}_l \mathbf{R}^H_l \right]_{f \left( n,\mathcal{B}_l \right)f \left( n,\mathcal{B}_l \right)} \leq P_n,\ \forall n \in \mathcal{B}_l, l \in \mathcal{L}.
\end{align}
\end{subequations}

Similar to the case without OBP, the AO algorithm described in Algorithm \ref{algo_1} can be used to solve $\mathcal{W}_{RS_{OBP}}^{ \left( S \right)} \left( P \right)$ such that the second stage precoders are optimised to achieve the best MMF rate for the system.

\section{Simulation Results}
\label{sec_simulation_result}
\begin{figure}
\centering
\begin{minipage}[t]{0.49\textwidth}
\centering
\includegraphics[width=\columnwidth]{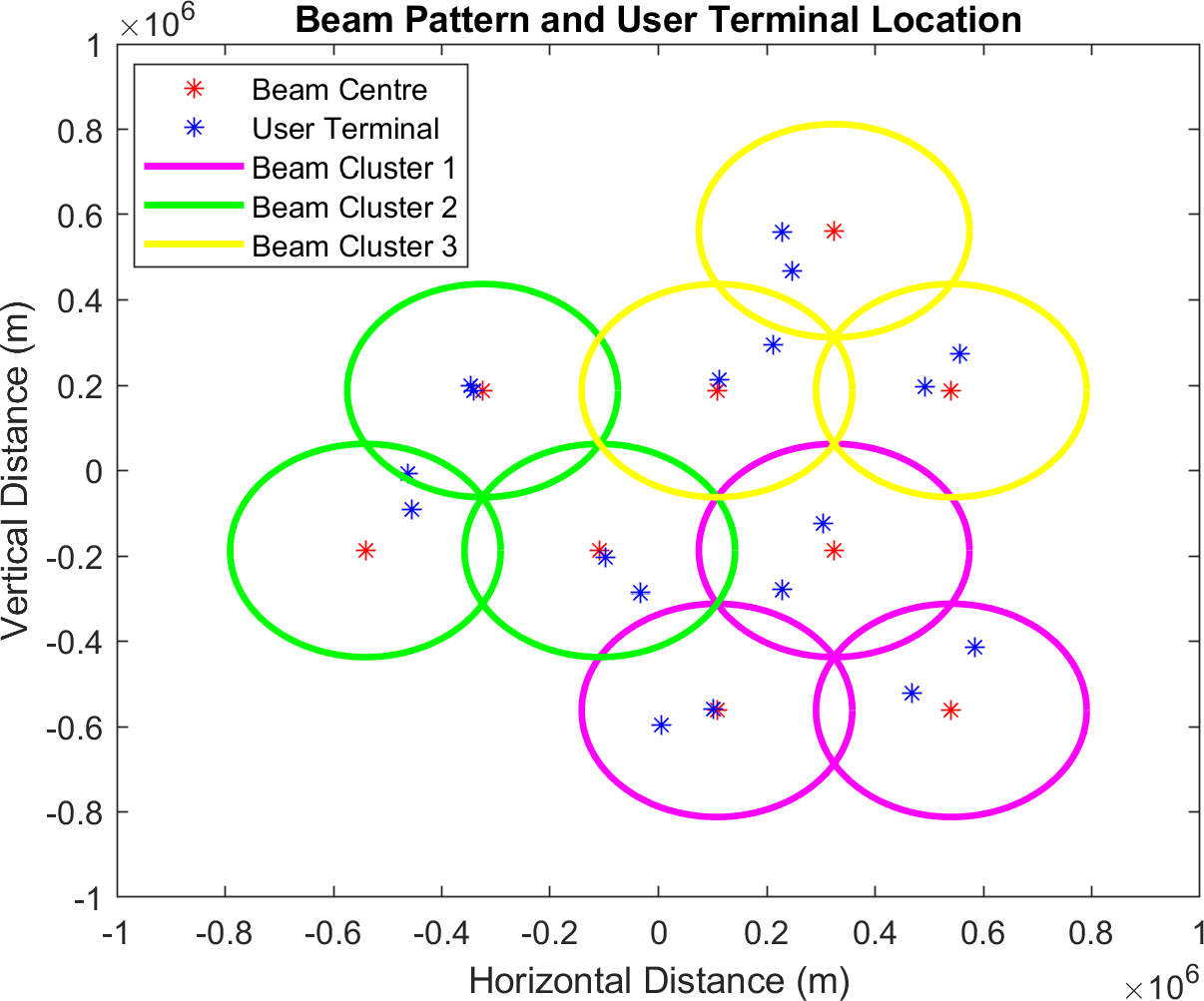}
\caption{Beam pattern of the three beam clusters with three beams each and user terminals uniformly distributed within each beam.}
\label{fig_topology}
\end{minipage}
\begin{minipage}[t]{0.49\textwidth}
\centering
\includegraphics[width=\columnwidth]{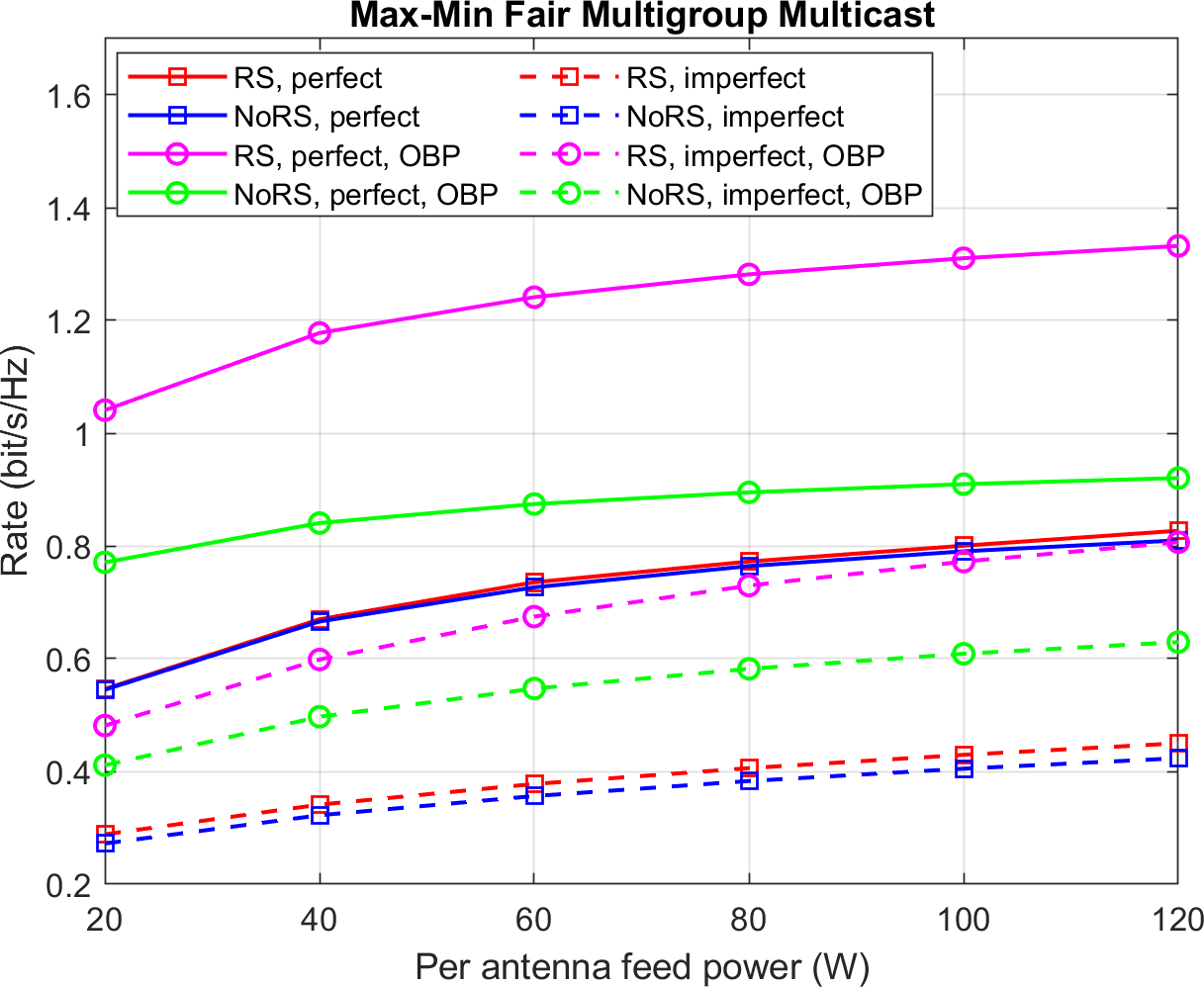}
\caption{Plot of MMF rate against per antenna feed power. $N = 9$ antenna feeds, $L = 3$ gateways, $\rho = 2$ users per group, imperfect CSIT: $\alpha = 0.6$, feeder link interference: $\delta = 0.8$.}
\label{fig_3_gateways}
\end{minipage}
\end{figure}

In this section, we evaluate the performance of RS and NoRS for multigateway multibeam satellite systems with feeder link interference under both perfect and imperfect CSIT. When forming the user link channels for the satellite system, the system parameters provided in Table \ref{table_1} are used \cite{yin2020rate}. There are $N = 9$ antenna feeds on the satellite and $L = 3$ gateways. Equal number of antenna feeds are used for every gateway therefore $B_l = B = 3$, $\forall l \in \mathcal{L}$. Within the $M = 9$ groups, there are $\rho = 2$ users per group, leading to a total of $K = 18$ users. The beam pattern and the user distribution can be seen in Fig. \ref{fig_topology}. Since cluster of antenna feeds are controlled by different gateways, the corresponding beams belonging to the same beam cluster are represented with the same colour.

For the feeder link channel between gateway $l$ and feeder link receiver $i$, we use the model $\mathbf{F}_{i,l} = q_l \mathbf{E}_{i,l}$ because it has been introduced in \cite{wang2019multicast}. $q_l$ is the atmospheric fading and it follows the same distribution as the atmospheric fading in the user link. Note that it does not change with the feeder link receiver because it is dependent on the gateway only. $\mathbf{E}_{i,l} = \mathbf{I}$ if $i=l$; otherwise, $\mathbf{E}_{i,l} = \delta_{i,l} \mathbf{1} \mathbf{1}^T$ where $\delta_{i,l}$ is the feeder link interference level. The model for $\mathbf{E}_{i,l}$ has been used in \cite{joroughi2016precoding} and \cite{joroughi2016feeder} to describe the feeder link channel gain as well as the inter-feeder link interference. In this paper, the feeder link interference levels are all set to be the same, i.e. $\delta_{i,l} = \delta = 0.8$, $\forall i,l \in \mathcal{L}$ \cite{wang2019multicast}. Despite using this specific model to simulate the results for this paper, other models for $\mathbf{F}_{i,l}$ are also applicable as part of the proposed framework.

\begin{table}
\caption{Satellite System Parameters}
\label{table_1}
\centering
\begin{tabular}{| c | c |} 
 \hline
 \textbf{Parameter} & \textbf{Value} \\
 \hline
 Frequency Band & Ka $\left (20\ \mathrm{GHz}  \right )$ \\
 Satellite Height & $35786\ \mathrm{km}\left ( \mathrm{GEO}\right )$ \\
 User Link Bandwidth & $500\ \mathrm{MHz}$ \\
 3 dB Angle & $0.4\degree$ \\
 Maximum Beam Gain & $52\ \mathrm{dBi}$ \\
 User Terminal Antenna Gain & $41.7\ \mathrm{dBi}$ \\
 System Noise Temperature & $517\ \mathrm{K}$ \\
 Rain Fading & $\left( \mu,\sigma \right) = \left( -3.125, 1.591 \right)$ \\
 \hline
\end{tabular}
\end{table}

When imperfect CSIT is considered, the channel estimates are formed by introducing errors to the actual user link channels and the feeder link channels, i.e. $\widehat{\mathbf{H}} = \mathbf{H} - \widetilde{\mathbf{H}}$ and $\widehat{\mathbf{F}} = \mathbf{F} - \widetilde{\mathbf{F}}$. The channel error vectors are i.i.d. and they are drawn from the complex Gaussian distribution, $\mathcal{CN} \left( 0,\sigma_e^2 \right)$ where $\sigma_e^2 = P^{-\alpha}$ \cite{yin2020rate}. $P$ is the total transmit power available at the satellite and the scaling factor, $\alpha$ is set to be 0.6. The sample size $S$ is 1000 when carrying out SAA for the deterministic problems. For a given channel estimate, the $s$-th conditional realisation of the channel is formed by adding back a sample of the channel error to the channel estimate, e.g. $\mathbf{H}^{\left( s \right)} = \widehat{\mathbf{H}} + \widetilde{\mathbf{H}}^{\left( s \right)}$, and the $s$-th sample of the channel error is drawn from the same error distribution. Note that $\mathbf{H}$ is the actual user link channel unknown to the gateways while $\mathbf{H}^{\left( s \right)}$ is part of the realisation set $\mathbb{H}^{\left( S \right)}$ available at the gateways used to calculate the SAFs.

For the per-antenna power constraint at the satellite, every antenna feed is allocated an equal amount of power, i.e. $P_n = P/N$, $\forall n \in \mathcal{N}$. We assume the noise power $\sigma_n^2$ to be 1 for simplicity. The tolerance level for convergence is set to be $10^{-4}$ and the solution to the optimisation problem is averaged over a total of 100 random channels to obtain the ergodic MMF rate. All optimisation problems were solved using the CVX toolbox \cite{grant2014cvx} on MATLAB.

Fig. \ref{fig_3_gateways} shows the MMF rates obtained after solving the optimisation problems from Section \ref{sec_feeder} and \ref{sec_obp} for a range of per antenna feed power at the satellite. With perfect CSIT, it can be seen that the performances using RS are always better than NoRS, despite only a slight gain when OBP is not used. When the CSIT quality drops, the performance gain for RS is still present, making it a more robust transmission scheme than NoRS. This solidifies the benefits of using the RS strategy for interference management in a multigateway multibeam satellite system, rather than just a terrestrial system. OBP is capable of minimising the effect of the feeder link interference and hence leading to higher MMF rates. Moreover, the gaps between RS and NoRS increase, which implies that the benefit of RS becomes more obvious when the feeder link interference is minimised through OBP. This is a desirable outcome to tackle the real practical problem of using multiple gateways. However, a more sophisticated satellite will need to be employed to allow for OBP payloads.

\begin{figure}
\centering
\begin{minipage}[t]{0.49\textwidth}
\centering
\includegraphics[width=\columnwidth]{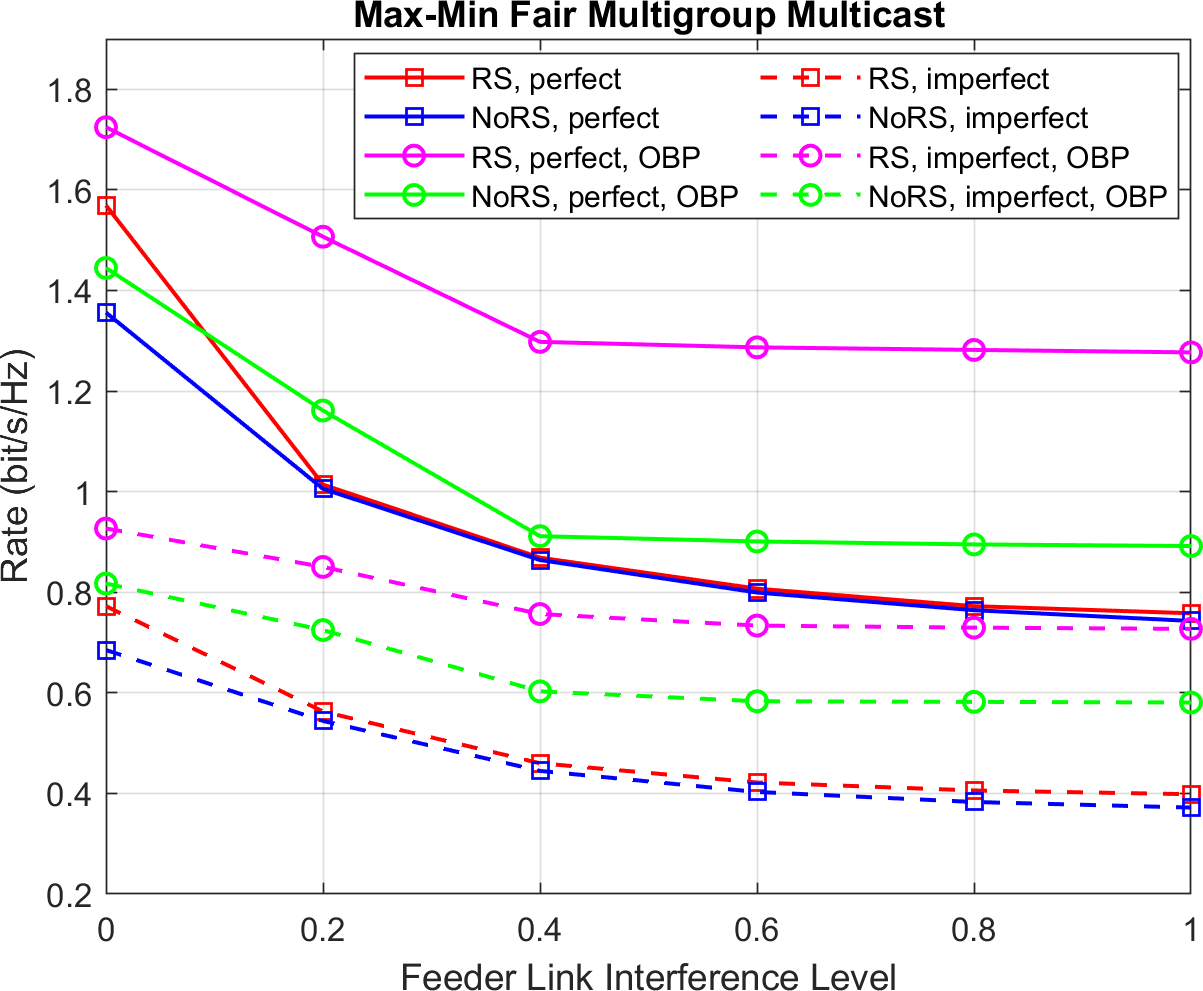}
\caption{Plot of MMF rate against feeder link interference level. $N = 9$ antenna feeds, $L = 3$ gateways, $\rho = 2$ users per group, $P/N = 80$ Watts, imperfect CSIT: $\alpha = 0.6$.}
\label{fig_interference}
\end{minipage}
\begin{minipage}[t]{0.49\textwidth}
\centering
\includegraphics[width=\columnwidth]{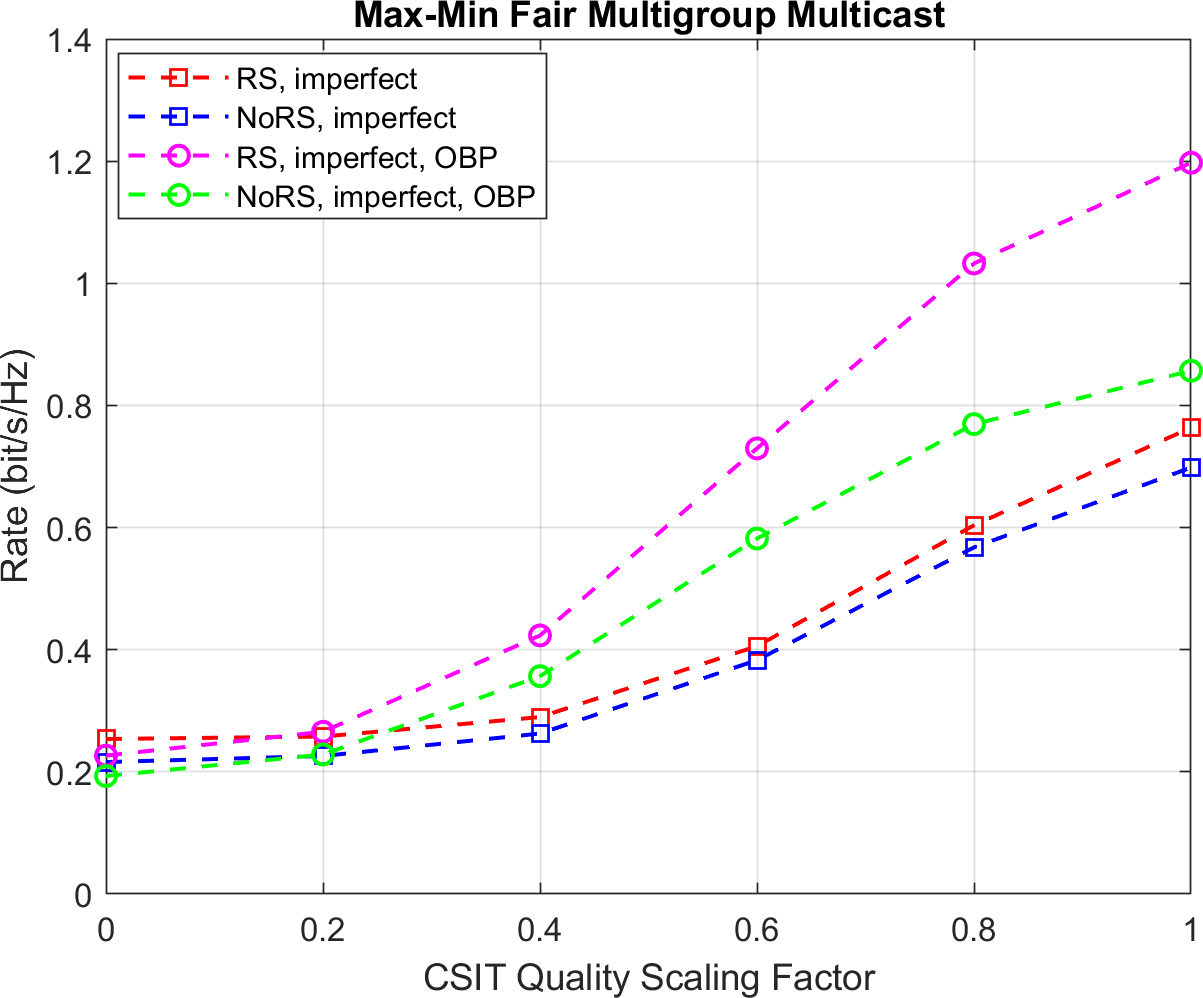}
\caption{Plot of MMF rate against CSIT quality scaling factor. $N = 9$ antenna feeds, $L = 3$ gateways, $\rho = 2$ users per group, $P/N = 80$ Watts, feeder link interference: $\delta = 0.8$.}
\label{fig_csit}
\end{minipage}
\end{figure}

In order to observe the effects of feeder link interference, the optimisation problems were solved once again but with varying feeder link interference levels, $\delta$. The per antenna feed power used was 80 Watts while other system parameters were kept the same. In Fig. \ref{fig_interference}, the use of OBP significantly improves the performance of the system under both perfect and imperfect CSIT. As the feeder link interference level increases, the MMF rates start to plateau, showing that feeder link interference would be the major issue in achieving higher rates. When $\delta = 0$, there is no interference between the gateways in the feeder link but the effect of atmospheric fading is still present. The gap between RS and NoRS is much bigger compared to others and this further implies that the benefit of RS is much significant when there is no feeder link interference. The use of OBP still provides higher rates for both RS and NoRS because the first stage precoder design would have dealt with the atmospheric fading in the feeder link.

As for the effect of imperfect CSIT, Fig. \ref{fig_csit} shows the various MMF rates achieved using different scaling factors, $\alpha$. It is expected to see that the MMF rates increase when the CSIT scaling factor increases since the CSIT quality becomes better. The performance of using RS is consistently better than NoRS at every scaling factor, making it a very robust transmission scheme when perfect CSIT is unobtainable. When $\alpha$ is large, the gain of RS over NoRS is more obvious when OBP is conducted. When $\alpha = 0$, the use of OBP performed worse than the case without OBP. This implies that the use of the channel realisation set for the feeder link channel is preferred over the effort of minimising the MSE of a channel with large CSIT errors.

\begin{figure}
\centering
\begin{minipage}[t]{0.49\textwidth}
\centering
\includegraphics[width=\columnwidth]{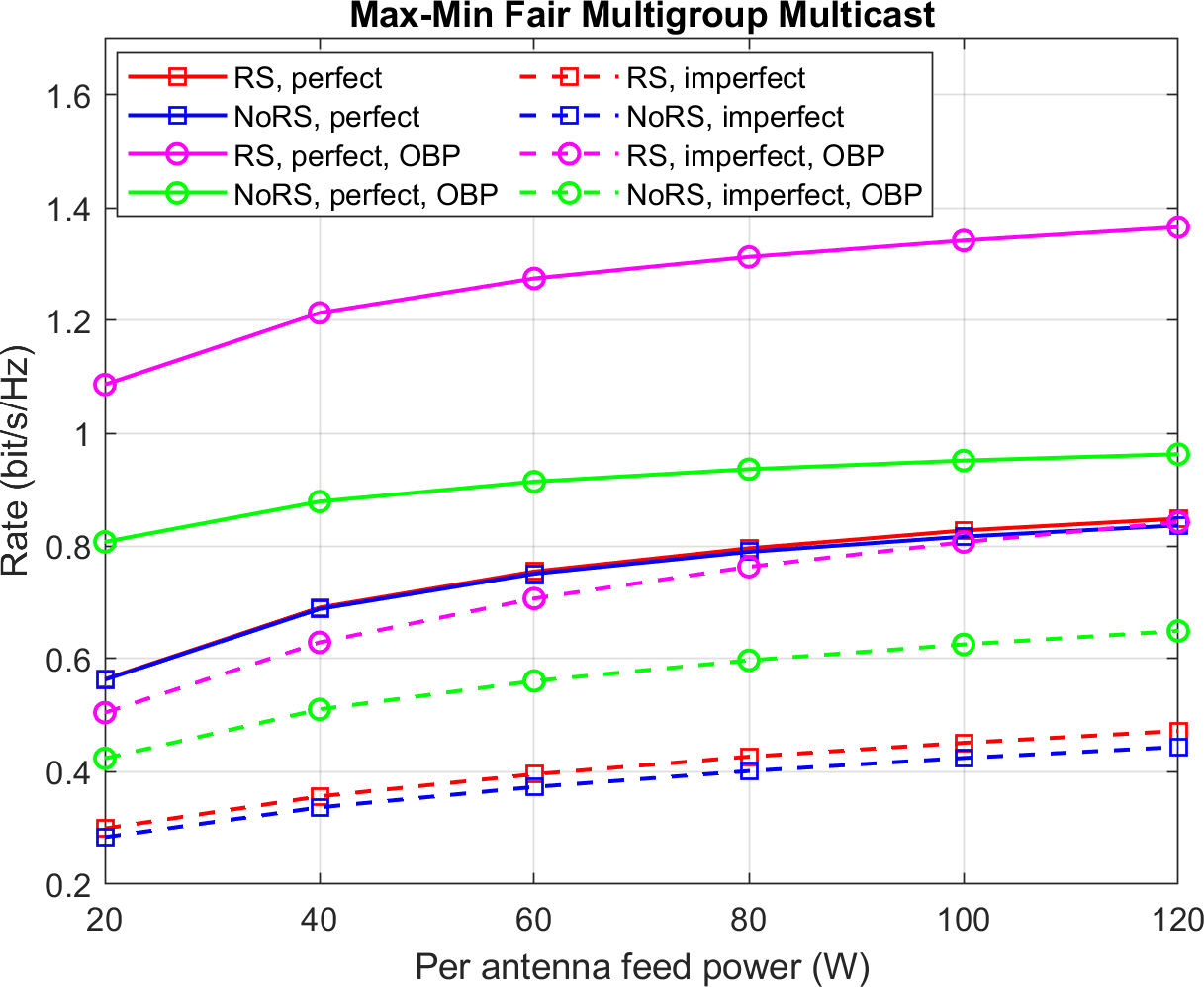}
\caption{Plot of MMF rate against per antenna feed power. $N = 9$ antenna feeds, $L = 3$ gateways, $G = \left[ 1,1,1,2,2,2,3,3,3 \right]$ users, imperfect CSIT: $\alpha = 0.6$, feeder link interference: $\delta = 0.8$.}
\label{fig_user_grouping}
\end{minipage}
\begin{minipage}[t]{0.49\textwidth}
\centering
\includegraphics[width=\columnwidth]{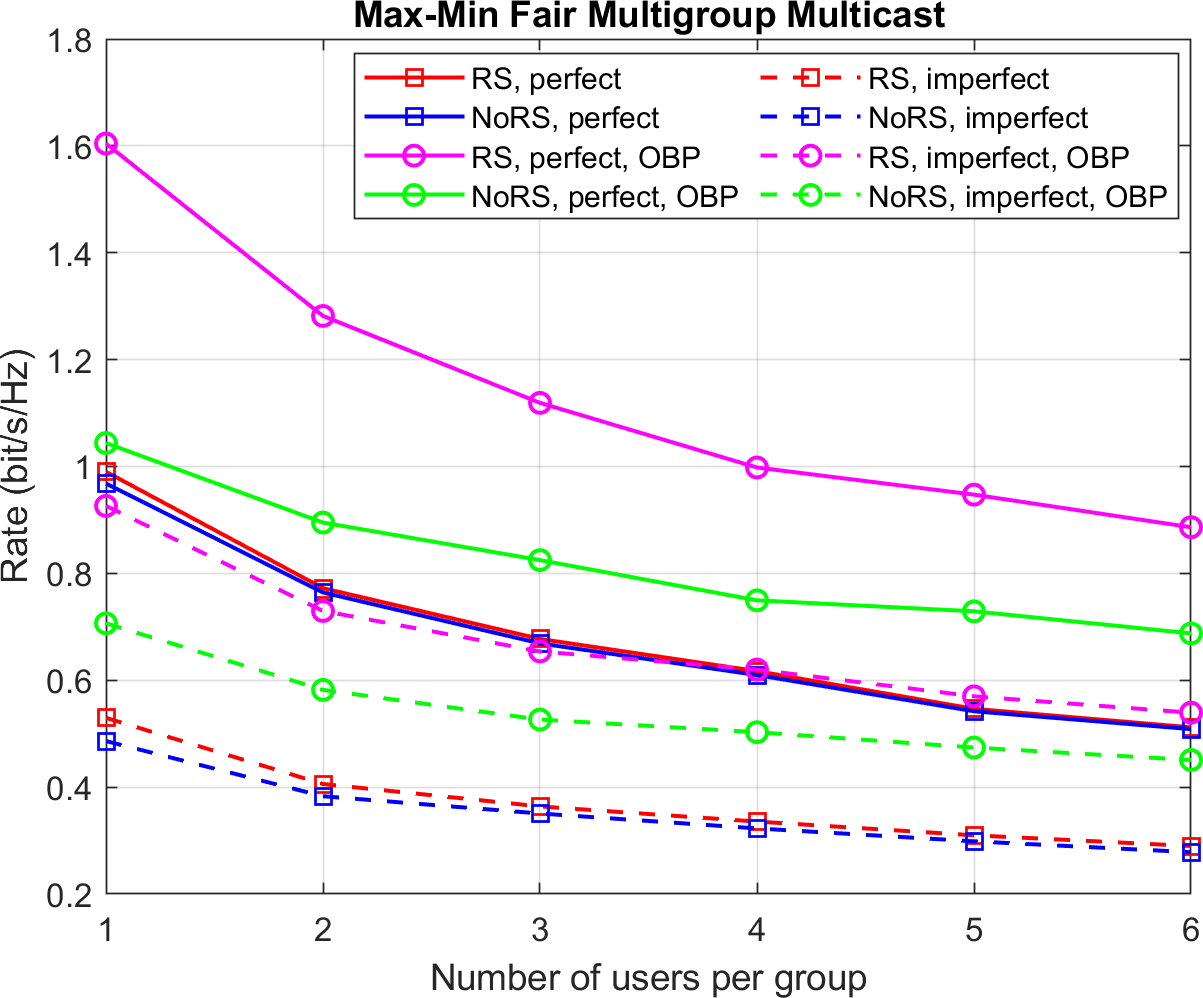}
\caption{Plot of MMF rate against number of users per group. $N = 9$ antenna feeds, $L = 3$ gateways, $P/N = 80$ Watts, imperfect CSIT: $\alpha = 0.6$, feeder link interference: $\delta = 0.8$.}
\label{fig_user_number}
\end{minipage}
\end{figure}

Rather than having equal number of users per group, the performance of the system using different number of users per group while keeping the same total number of users was looked into. Based on Fig. \ref{fig_user_grouping}, the benefits of using RS and OBP are still noticeable. When compared to the uniform setting, the custom setting led to a very slight increase in the MMF rate for all cases despite the change in the number of users in certain groups.

Finally, the effect of number of users per group, $\rho$ on the performance of the system was investigated. Fig. \ref{fig_user_number} depicts that the MMF rate decreases when the number of users per group increases for all cases. This is because every group has only one precoder for its private stream, hence the users within a group will need to share this precoder even though they all have different channels. As a result, the user with the worst SINR would then influence its group rate dramatically. Despite this performance degradation, RS is still capable of providing some gains as compared to NoRS and the use of OBP still results in higher MMF rates.

\section{Conclusion}
In this paper, we propose the use of the RSMA to manage the interference contributed by the feeder link and the user link in a multigateway multibeam satellite communication system. The MMF problem is formulated and the equivalent WMMSE problem is solved using the AO algorithm for both perfect and imperfect CSIT cases. Two-stage precoding using OBP is also implemented such that the feeder link interference can be managed before the user link interference. The use of RS at each gateway is highly recommended as it is able to handle intra-gateway interference effectively and cope with CSIT uncertainty when compared to the conventional method of using NoRS. Finally, if resources allow, OBP should be utilised to enlarge the benefits provided by RSMA and further improve the system performance.

\ifCLASSOPTIONcaptionsoff
  \newpage
\fi

\bibliographystyle{IEEEtran}
\bibliography{ref}
\end{document}